\documentclass[
 reprint,
 amsmath,amssymb,
 aps,
 superscriptaddress,
]{revtex4-1}

\usepackage[all,defaultlines=2]{nowidow} 

\usepackage[section]{placeins} 
\usepackage{color} 
\usepackage{cancel} 
\usepackage{lineno} 
\usepackage[utf8]{inputenc} 
\usepackage{natbib} 
\usepackage{csquotes} 
\usepackage[caption = false]{subfig} 
\usepackage{hyperref} 
\hypersetup{colorlinks=true, linkcolor=blue, filecolor=magenta, urlcolor=blue}  

\usepackage{graphicx}
\usepackage{bm}
\usepackage{textcomp}
\usepackage{upgreek}

\begin{document}

\title{How heat controls fracture: the thermodynamics of creeping and avalanching cracks}

\author{Tom Vincent-Dospital}
\email{vincentdospitalt@unistra.fr}
\affiliation{Université de Strasbourg, CNRS, Institut de Physique du Globe de Strasbourg, UMR 7516, F-67000 Strasbourg, France}
\affiliation{SFF Porelab, The Njord Centre, Department of physics, University of Oslo, N-0316 Oslo, Norway}

\author{Renaud Toussaint}
\email{renaud.toussaint@unistra.fr}
\affiliation{Université de Strasbourg, CNRS, Institut de Physique du Globe de Strasbourg, UMR 7516, F-67000 Strasbourg, France}
\affiliation{SFF Porelab, The Njord Centre, Department of physics, University of Oslo, N-0316 Oslo, Norway}

\author{Stéphane Santucci}
\affiliation{Université de Lyon, ENS de Lyon, Université Claude Bernard, CNRS, Laboratoire de Physique, F-69342 Lyon, France}
\affiliation{Mechanics of disordered media laboratory, Lavrentyev Institute of Hydrodynamics of the Russian Academy of Science}

\author{Loïc Vanel}
\affiliation{Université de Lyon, Université Claude Bernard, CNRS, Institut Lumière Matière, F-69622 Villeurbanne, France}

\author{Daniel Bonamy}
\affiliation{Universit{\'e} Paris-Saclay, CNRS, CEA Saclay, Service de Physique de l’Etat Condens{\'e}, F-91191 Gif-sur-Yvette, France}

\author{Lamine Hattali}
\affiliation{Université Paris-Sud, CNRS, Laboratoire FAST, UMR 7608, F-91405 Orsay, France}

\author{Alain Cochard}
\affiliation{Université de Strasbourg, CNRS, Institut de Physique du Globe de Strasbourg, UMR 7516, F-67000 Strasbourg, France}

\author{Eirik G. Flekk\o y}
\affiliation{SFF Porelab, The Njord Centre, Department of physics, University of Oslo, N-0316 Oslo, Norway}

\author{Knut J\o rgen M\aa l\o y}
\affiliation{SFF Porelab, The Njord Centre, Department of physics, University of Oslo, N-0316 Oslo, Norway}


\begin{abstract}
          While of paramount importance in material science, the dynamics of cracks still lacks a complete physical explanation. The transition from their slow creep behavior to a fast propagation regime is a notable key, as it leads to full material failure if the size of a fast avalanche reaches that of the system. We here show that a simple thermodynamics approach can actually account for such complex crack dynamics, and in particular for the non-monotonic force-velocity curves commonly observed in mechanical tests on various materials. We consider a thermally activated failure process that is coupled with the production and the diffusion of heat at the fracture tip. In this framework, the rise in temperature only affects the sub-critical crack dynamics and not the mechanical properties of the material. We show that this description can quantitatively reproduce the rupture of two different polymeric materials (namely, the mode I opening of polymethylmethacrylate (PMMA) plates, and the peeling of pressure sensitive adhesive (PSA) tapes), from the very slow to the very fast fracturing regimes, over seven to nine decades of crack propagation velocities. In particular, the fastest regime is obtained with an increase of temperature of thousands of kelvins, on the molecular scale around the crack tip. Although surprising, such an extreme temperature is actually consistent with different experimental observations that accompany the fast propagation of cracks, namely, fractoluminescence (i.e.,\,the emission of visible light during rupture) and  a complex morphology of post-mortem fracture surfaces, which could be due to the sublimation of bubbles.
\end{abstract}
                           
\maketitle


\section{Introduction}

The rupture of solids is often described by empirical observations rather than by fully understood physical models. One of the earliest formalisms is that by Griffith in 1921\,\cite{Griffith1921}: the propagation of cracks is described as a threshold phenomenon, only obtained when loading their encompassing matrix above a critical fracture energy. To the first order, this view matches the behavior of brittle bodies, which suddenly snap passed a certain elastic deformation. Analytical models of cracks propagating in lattices suggested\,\cite{Slepyan_81,Kulakhmetova_84,Slepyan_Marder,Marder1995} that such an instability arises from the discrete nature of matter at the atomic scale. Indeed, these models revealed a minimum propagation velocity, comparable to that of the mechanical waves in the considered material, above which the advance of a fracture tip through the network of molecular bonds can be self maintained by the emission of high frequency phonons. There, the energy binding two lattice nodes is defined as a covalence-like barrier\,\cite{covalence}.
While this description\,\cite{Slepyan_81} does not allow for slow propagation, it is acknowledged that a crack loaded well below the fast rupture threshold is still growing, but at creeping rates that are orders of magnitude below that of a `dynamic' fracture (e.g.,\,\cite{Brenner,Zhurkov1984}). An approach to explain such a creep in a way that is compatible with Griffith's formalism\,\cite{Griffith1921} is to consider that the fracture energy is not an intrinsic material property, but is instead a particular function of the propagation velocity\,(e.g.,\,\cite{PMMA_Doll}). One hence simply obtains a lower crack speed if providing a lesser mechanical load. Alternatively, the creep regime is well modelled\,\cite{Brenner, Zhurkov1984, crack_nucleation_rate, Pomeau, roux, Santucci2004, Vanel_2009, Lengline2011, Cochard} by thermally activated sub-critical laws such as Arrhenius-like growth rates (e.g.,\,\cite{kinetics}), and thermodynamics has thus emerged as a framework to describe the slow failure. In such descriptions, that are sometimes referred to as `stress corrosion', a variation of fracture energy with velocity is not particularly called for, as the molecular agitation allows the crack to progress at loads below an intrinsic rupture threshold.\\
In practice, and depending on the material being broken, both the slow and the fast propagation regimes can be observed for a same range of applied loads\,\cite{Marshall_1974,tape2}. A hysteresis holds and the growth rate of a fracture is then depending on the actual mechanical history, rather than only on the instantaneous mechanical load. Maugis and Barquins\,\cite{MaugisBarquins_1978,Maugis1985} early suggested that the description of the slow and the fast regimes, as well as that of the hysteresis, could be qualitatively unified by reinterpreting Griffith's criteria\,\cite{Griffith1921}, if one could account for the temperature and velocity dependent viscoplasticity that occurs around crack tips\,\cite{orowan1954, Irwin1957}. More specifically, \citet{Marshall_1974} and then Carbone and Persson\,\cite{carbonePersson, carbonePersson_long} proposed that the induced heat associated to such a plasticity might locally soften the matter around a crack and that some thermal weakening (i.e., the abrupt transition from slow creep to fast failure due to a thermal process) arises from the related reduction of the material elastic moduli.\\
In this work, we propose a quantitative unifying model of the two propagation regimes that disregards such a softening effect, hence stating that some variations in the material mechanical properties are not necessarily required to obtain a slow-to-fast-crack transition. We focus instead on how the thermal dissipation, and the subsequent rise in tip temperature, affect the front sub-critical growth, as understood by statistical physics and an Arrhenius-like law. In some previous works, we indeed studied how such sub-critical laws, at fixed room temperature, well describe creep; in fibrous and polymeric materials (namely, paper sheets and polymethylmethacrylate, PMMA), they notably account for the mean kinetics of slow rupture fronts under various loading conditions\,\cite{Santucci_2006,Lengline2011,Vanel_2009}. When, in addition, taking into account these media structure and heterogeneities in fracture energy, such sub-critical laws also reproduce the intermittent dynamics of failure; in particular, the size distribution of crack jumps\,\cite{Santucci2004} and the front roughening properties\,\cite{Cochard}. Here we neglect any spatial variation of the fracture energy, but let the crack tip temperature vary as a function of the front velocity and of the applied mechanical load. Indeed, in a previous experimental and theoretical study of the tearing-induced heating in paper sheets\,\cite{ToussaintSoft}, we were able to relate the temperature field around moving cracks to a certain percentage of the mechanical energy which gets converted into heat as the tip advances. More recently, this rise in temperature was fed back into a sub-critical growth law and showed\,\cite{TVD1} that one can thus obtain a dynamics model holding numerous qualitative similarities with the observed behavior of cracks, namely, two stable phases of propagation and a critical point that is similar to a brittle-ductile transition (e.g.,\,\cite{Scholz1988}). Here, this model is first reintroduced (section\,\ref{section:model}) and then shown to quantitatively capture the fracturing dynamics of two different polymeric materials, over the full range of velocities (section\,\ref{section:fits}), namely, acrylic glass (PMMA) and pressure sensitive adhesives (PSA). In both these media, some extensive experimental work has been carried out by different groups to quantify the two rupture regimes (e.g., see Refs.\,\cite{temp_PMMA, PMMA_Doll, Fineberg1991, PMMA1, PMMA2, Maloy2006, Tallakstad2011} for PMMA and Refs.\,\cite{tape2, fibril, tape1, stephane_tape_PRL} for PSA) and our proposed model accounts for the experimental curves of applied load versus crack velocity, from the slowest (micrometers per second) cracks to the fastest (hundreds of meters per second) ones. Such a match suggests that the growth of cracks could be sub-critical (i.e., as stated by the model) over a far wider velocity range than what is commonly accepted, that is, even at propagation velocities approaching that of mechanical waves. Indeed, we infer that the load threshold at which cracks typically shift to the fast phase is actually smaller than the intrinsic rupture energy, as a result from the boosted thermal activation around the front. In particular, we predict that crack tips can reach thousands of degrees on the molecular scale (i.e., over a few atoms around the front), when they quickly avalanche. Although such high temperatures are today rarely considered, they have long been proposed (e.g.,\,\citet{ricelevy1969}), and we here discuss (section\,\ref{concl}) how they are inline with several observables that sometimes accompany the fast propagation of cracks, namely, the emission of visible light at their tips (i.e., fractoluminescence\,\cite{WEICHERT_1978, tribo_blackbody, Bouchaud2012}) and the existence of bubbles on their postmortem surfaces, that can nucleate secondary rupture fronts\,\cite{Smekal1953,RaviChandar1997}.

\section{From thermal creeping to thermal weakening \label{section:model}}

\subsection{The kinetics of sub-critical rupture}

We here consider a refinement of the propagation model already introduced by \citet{TVD1}, that did not compare it to any actual, experimental, crack propagation. Let us start by restating the various components of this model.\\
We consider the velocity $V$ of cracks to be ruled by the competition, at their tips, between breaking and healing processes \cite{RICE1978} (or see Ref.\cite{lawn_1993}, chpt 5.5.1). As many authors before us (e.g.,\cite{Brenner, Zhurkov1984, lawn_1993}), we propose that these processes are, at least in part, sub-critical, and are governed by some Arrhenius-type laws (e.g.,\cite{kinetics}, chpt. 1.8.1). The activation energies of these laws are thus exceeded by the thermal bath according to a probabilistic Boltzmann distribution\cite{kinetics}. The rupture activation energy can then be written as $(U_c-U)$: the difference between the mechanical energy $U$ that is stored in the tip bond and a critical rupture energy $U_c$, at which this bond fails. The latter should typically be comparable to a few electronvolts, which is a standard value for atomic covalence (e.g., see appx. E in Ref.\cite{covalence}). Of course, depending on the studied material, $U_c$ could also be dominated by the typically weaker binding energies of hydrogen or Van der Waals bonds, and its actual value may thus lie within a few orders of magnitude. In any case, as we are here introducing a mesoscopic law for the rupture dynamics (i.e., an Arrhenius growth), $U_c$ should be understood as a mean material property, representative of the various strengths of the links that break along a crack course. Such a statistical definition will also apply to most of the parameters that we will henceforward consider.
Similarly to the rupture barrier, the activation energy to heal the atomic connections can be written as $(U_h+U)$. There, $U_h$ is an intrinsic repulsive energy barrier that two atoms need overcome to bond, in addition to which the thermal bath at the healing link also needs to compensate for the applied stretch $U$ of the tip. With these considerations, the propagation velocity of a crack is then modelled by
\begin{equation}
   V = \nu d_0 \exp \left( {-\frac{U_c-U}{k_B T}} \right) - \nu d_0 \exp \left( {-\frac{U_h+U}{k_B T}} \right),
   \label{arrh00}
\end{equation}
where the first term is the forward rupture velocity of the crack and the second one is the backward healing velocity. In this equation, we denote $d_0 \sim 2$\,{\AA} the inter-atomic distance, $k_B$ is Boltzmann's constant $\sim 1.38\,\times\,10^{-23}$ m\textsuperscript{2}\,kg\,s\textsuperscript{-2}\,K\textsuperscript{-1}, $T$ is the absolute temperature at the crack tip and $\nu$ is the collision frequency in the molecular bath (e.g.,\,\cite{kinetics}, chpt. 4.1). Each exponential in Eq.\,(\ref{arrh00}) is a probability term (i.e., the probability, challenged every $1/\nu$ second, that the thermal bath exceeds one of the activation energies and that the crack hence advances or retreats by a step $d_0$). As such, these terms cannot be greater than $1$ and, while the healing one always meets this condition, $U \geq U_c$ corresponds to an over-critical propagation regime where
\begin{equation}
   V = \nu d_0 \Bigg[ 1 - \exp \left( {-\frac{U_h+U}{k_B T}} \right) \Bigg].
   \label{arrh02}
\end{equation}
\begin{figure}[t]
  \includegraphics[width=1\linewidth]{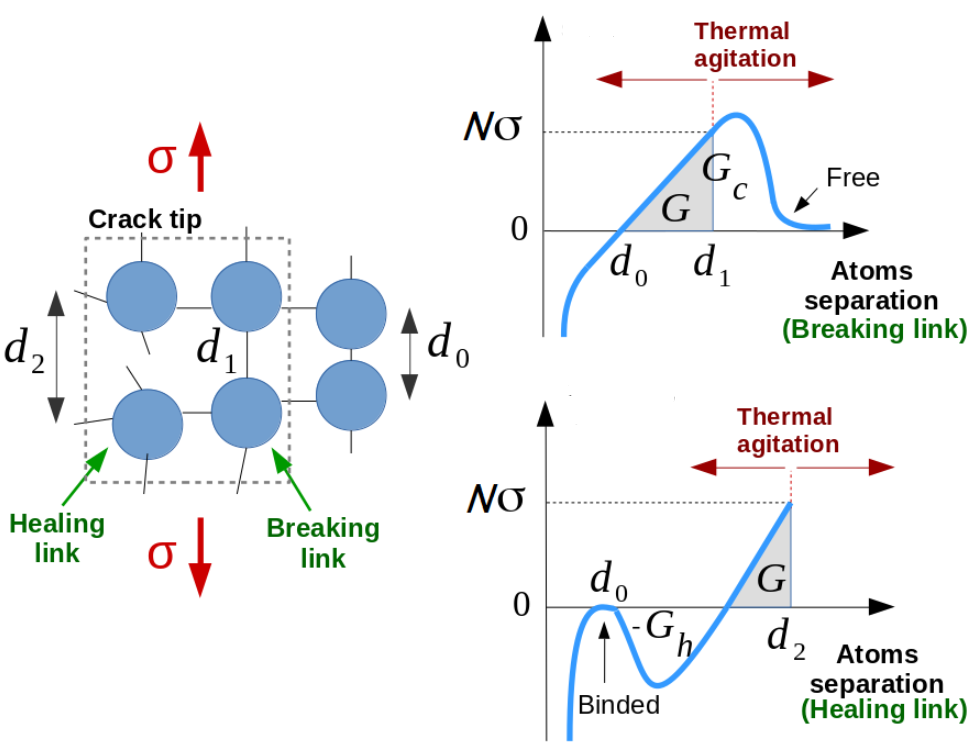}
  \caption{(Left): simplified atomic view of the breaking/healing site at the crack tip. (Top right): Generic tip stress $\sigma$ normalised by the stress shielding factor $N$ versus atom separation for the active breaking link. (Bottom right): Generic tip stress normalised by the stress shielding factor versus atom separation for the active healing link. The grey areas are the energy release rate $G$. At this load, $d_1$ and $d_2$ are the mean extensions of, respectively, the breaking and the healing link, while $d_0$ is the unstressed atom separation. On the breaking link graph: the area below the curve for $d>d_0$ is the intrinsic surface fracture energy $G_c$. The thermal agitation may overcome the remaining $G_c-G$ barrier. Although the healing link is initially broken, an energy input is required to move the two particles closer to each other, due to the neighbouring unbroken links stretched at a load $G$. In addition, when the atoms separation gets smaller, the thermal agitation also needs to overcome a repulsive energy barrier $G_h$ (the area below the atoms separation axis in this figure) before reforming the bond.}
  \label{fig:schem}
\end{figure}
The product $\nu d_0$ is a maximal velocity, that we will further denote $V_0$, at which a fracture front can advance, when its tip atomic bonds snap each time they are challenged and never heal. In theory\,\cite{kinetics}, the frequency $\nu$ is temperature dependent, with $V_0\sim\sqrt{k_B T/m}$ where $m$ is the mass of an atom or a molecule, but this dependence is small compared to that of the neighbouring exponential terms, so that we here neglect it. In our context of rupture kinetics, and more practically, it was notably proposed\,\cite{Freund1972,Marder1995} that such a nominal velocity $V_0$ is in the order of that of the medium Rayleigh waves, as quicker fractures then propagate in a specific supersonic regime\,\cite{supersonic1, supersonic2}, which is not here considered.\\
In our description, $U$ is the physical quantity that describes the load of a crack on the microscopic level, and that governs most of its dynamics. However, at the lab scale, $U$ is not a measurable quantity. The energetic level at which a crack progresses is rather characterized  by the macroscopic energy release rate $G$, which is the amount of energy that a fracture dissipates to grow by a given unit of measurable area\,\cite{Griffith1921}. This energy dissipation may be of diverse nature, and is to cause a relative reduction in potential energy near the tip. We will denote $N>1$ the factor for this reduction, so that $U\sim d_0^2G/N$. More commonly, mechanical shielding is described with the introduction of a plastic process zone of radius $\xi$ around the crack front, where the dissipation occurs. To follow this canonical framework, we define a radius $\xi$ that is relative to the length of an atom link, such that $2\xi/d_0=N$. The intensity of the mechanical shielding (i.e., the relation between the potential energy $U$ stored in the rupturing bond and the macroscopic energy dissipation $G$) then writes as
\begin{equation}
    U\sim\frac{d_0^3G}{2\xi},
    \label{UG2}
\end{equation}
By additionally introducing $G_c=2\xi U_c/d_0^3$ and $G_h=2\xi U_h/d_0^3$, the respective equivalents in the energy release rate framework of $U_c$ and $U_h$, one can re-write Eqs.\,(\ref{arrh00}) and (\ref{arrh02}) as functions of $G$:
\begin{equation}
\resizebox{1\linewidth}{!}{$
\begin{aligned}
   V = V_0 \Bigg[ \exp \left( {-\frac{d_0^3(G_c-G)}{2\xi k_B (T_0+\Delta T)}} \right) - \exp \left( {-\frac{d_0^3(G_h+G)}{2\xi k_B (T_0+\Delta T)}} \right)& \Bigg]\\
   \text{ when } G<G_c&\\
   V = V_0 \Bigg[ 1 - \exp \left( {-\frac{d_0^3(G_h+G)}{2\xi k_B (T_0+\Delta T)}} \right) \Bigg] \hspace{1cm} \text{ when } G \geq G_c&.
\end{aligned}
$}
   \label{arrh1}
\end{equation}
We have here also written $T$ as $T_0+\Delta T$, where $T_0$ is the absolute room temperature ($\sim296$\,K) and $\Delta T$ is any deviation from this background value, as we will proceed to propose that the tip temperature can vary.
\\Note finally that one could also write this relation as a function of the mechanical stress $\sigma$ that is applied at the crack tip, using $G(d)=\int_{d_0}^{d} N\sigma(d') \mathrm{d} d'$, where $d_0$ is the nominal separation of atoms in an unloaded matrix (i.e., at $G=0$) and $d$ is the actual atom separation at the crack tip. Figure\,\ref{fig:schem} illustrates such a link between $G$ and $\sigma$ and summarizes, in a simplified atomistic view, how the thermal bath allows to overcome the surface energy barriers for breaking and healing atomic bonds, $G_c-G$ and $G_h+G$, as per Eq.\,(\ref{arrh1}).

\subsection{Heat dissipation and tip temperature rise}

\begin{figure}[b]
  \includegraphics[width=1\linewidth]{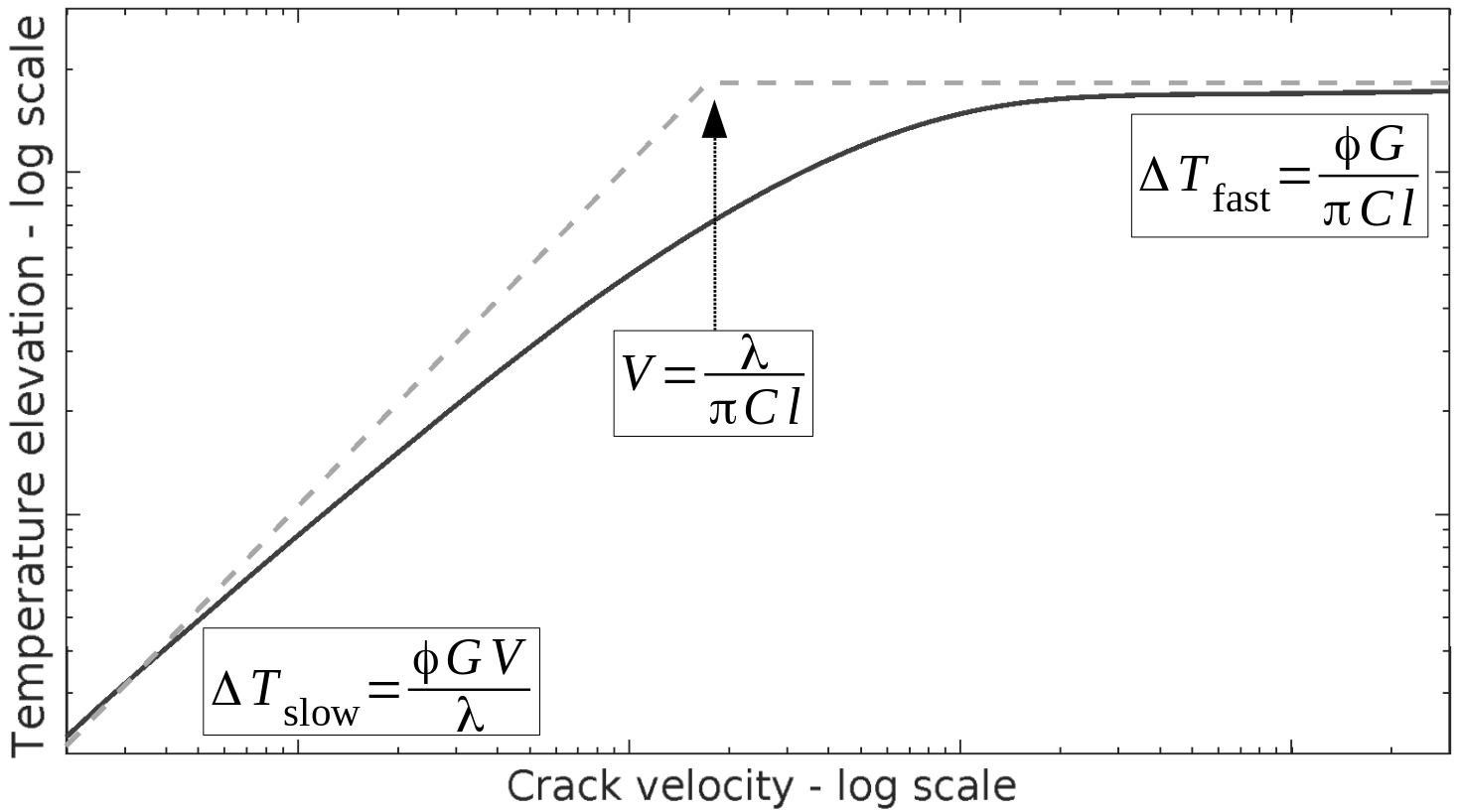}
  \caption{Steady thermal elevation at a crack tip for various propagation velocities, due to the diffusion equation (\ref{tempdiff}) (plain plot). The approximations ${\Delta T}_{\text{fast}}$ and ${\Delta T}_{\text{slow}}$, from Eqs.\,(\ref{slow}) and (\ref{fast}), are shown for comparison (dotted plots). The axes are not annotated for the sake of generality.}
  \label{fig:DTV}
\end{figure}
In the model we have introduced, one needs to further account for the energy which is dissipated around the running tip ($G$), as, even if it is mechanically lost, we will here show that it can maintain a strong effect on the crack dynamics. While the energy dissipation can be of several forms, ranging from the emission of mechanical waves\,\cite{crackwaves} damped in the far field, to the nucleation of defaults in the matrix\,\cite{rice_surface} (i.e., crazing\,\cite{PMMA_Doll, Crazing}), we here focus on the release of heat around the fracture tip\,\cite{WEICHERT_1978, ToussaintSoft}. We thus call $\phi$ the percentage of $G$ that is converted into some local rise in internal energy, and hence in temperature, and denote $l$ the typical size over which this process occurs.
As the heat, released on a production zone of area $\pi l^2$ close to the tip, is to diffuse in the whole bulk, the resulting temperature elevation $\Delta T$ can be modelled (e.g.,\,\cite{ToussaintSoft}) by the standard diffusion equation:
\begin{equation}
   \frac{\partial (\Delta T)}{\partial t} = \frac{\lambda}{C} \nabla^2 (\Delta T) + \frac{\phi G V}{C \pi l^2}f,
   \label{tempdiff}
\end{equation}
where $\lambda$ is the medium's thermal conductivity, and $C$ is the volumetric heat capacity. The last term of this equation is a source term only valid in the heat production zone. The support function $f$ of this zone is $1$ inside of it and $0$ otherwise, and the thermal source term is proportional to $\phi G V$, that is the dissipated power per unit of crack length deposited in the advancing zone. Although governed by Eq.\,(\ref{tempdiff}), $\Delta T$ at the rupture front can approximate to far simpler expressions. It was indeed shown\,\cite{ToussaintSoft} that, at low propagation velocities, the temperature elevation at the centre of the heat production zone (i.e., the crack tip) is only governed by the diffusion skin depth $\delta=\sqrt{\lambda\tau/(\pi C)}$ upon the passage of the production zone of extension $l$ within the time $\tau=l/V$. For fast cracks however, when $\delta$ becomes smaller than $l$, the generated heat can barely diffuse out of its source zone and $\Delta T$ is then constrained by $l$. We thus have
\begin{equation}
{\Delta T}_{\text{slow}} \sim \frac{\phi GV\tau}{C(\pi\delta^2)} = \frac{\phi GV}{\lambda},
\label{slow}
\end{equation}
\begin{equation}
{\Delta T}_{\text{fast}} \sim \frac{\phi GV\tau}{C(\pi l^2)} = \frac{\phi G}{\pi C l}.
\label{fast}
\end{equation}
Figure \ref{fig:DTV} shows the general evolution of $\Delta T$ at the tip with $V$, according to Eq.\,(\ref{tempdiff}) solved by numerically integrating the heat diffusion kernel\,\cite{HeatCond}. Note that $\Delta T$ in Eq.\,(\ref{tempdiff}) is a temperature field as shown for instance in the inset of Fig.\,\ref{fig:altuglas}, but we are here mainly interested in its value at the centre of the heat production zone (i.e., where the rupture process occurs). Figure \ref{fig:DTV} also shows how the two expressions of Eqs.\,(\ref{slow}) and (\ref{fast}) approximate for the tip temperature.

\subsection{Model phase behavior}

\begin{figure*}
  \includegraphics[width=1\textwidth]{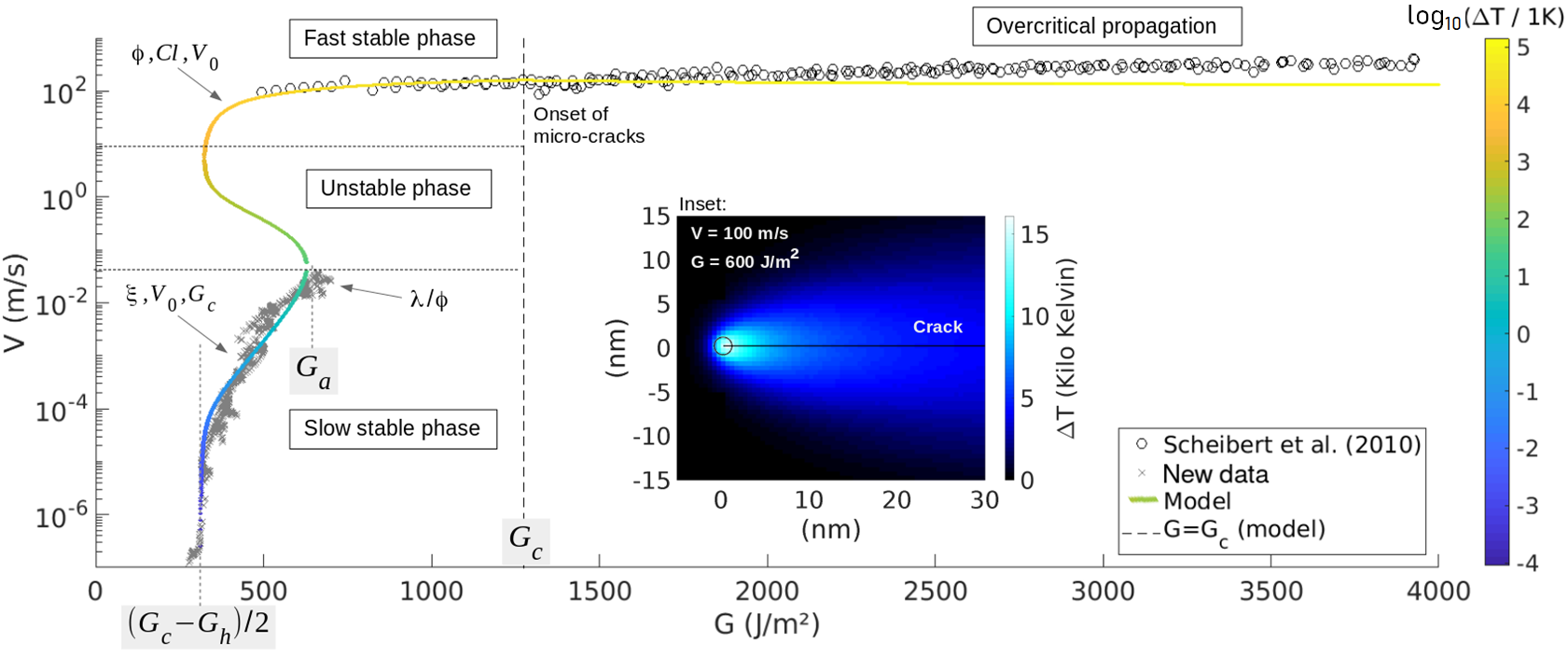}
  \caption{Crack velocity $V$ as a function of the energy release rate $G$ as predicted by Eqs.\,(\ref{arrh1}) and\,(\ref{tempdiff}) (plain curve) and fitted to the PMMA experimental data\,\cite{PMMA1}. The arrows indicate to which model parameters each part of the curve is mainly sensitive, and the main color scale specifies at which temperature the crack tip is modelled to be. The load $G_a$ is an avalanche threshold beyond which a front can only propagate quickly and $G_c$ is the modelled microscopic energy barrier for rupture. Below the asymptote at $(G_c-G_h)/2$, fronts cannot propagate forward due to some dominating healing processes. The inset shows, for a given point of the curve: $V \sim 100$\,m\,s\textsuperscript{-1} and $G \sim 600$\,J\,m\textsuperscript{-2}, the associated modelled temperature field around the front. For readability, the color map is there different from the main one, and the circle corresponds to the tip of radius $l$, where the extra heat is emitted. $\Delta T$ of the main model curve is the value at the centre of the circle. At loads beyond $G=G_c$, micro-cracks begin to nucleate\,\cite{PMMA1}, as shown further in Fig \ref{fig:micro}, which shows a zoom of the fast branch.}
  \label{fig:altuglas}
\end{figure*}

We have now derived the two constitutive equations of our fracture dynamics model: Eq.\,(\ref{arrh1}), that gives the velocity of a crack as a function of its tip temperature, and Eq.\,(\ref{tempdiff}), that governs the thermal state around a progressing front. In a previous work\,\cite{TVD1}, we have simultaneously solved these two equations and, focusing on their steady state, showed that they predict two stable phases for the propagation of cracks. These two behaviors are shown by the plain curve in Fig.\,\ref{fig:altuglas}, and are there labelled `Slow stable phase' and `Fast stable phase'. The first one, as its name suggests, is a slow one, where $\Delta T$ stays small compared to $T_0$, such that the growth rate is mainly governed by the medium fracture energy $G_c$ (i.e., as indicated by Eq.\,(\ref{arrh1})). This slow branch ceases to exist beyond a particular load $G=G_a$. The second phase is reached when the generated heat (and hence $\Delta T$) significantly overcomes the background temperature. From the Arrhenius law (\ref{arrh1}), the growth rate then significantly increases, so that the crack is said to be thermally weakened. Note, in Fig.\,\ref{fig:altuglas}, how both phases coexist for a certain range of energy release rates: a hysteresis situation holds (e.g., between $G=300$\,J\,m\textsuperscript{-2} and $G=G_a$ in Fig.\,\ref{fig:altuglas}). When this is the case, the model also predicts\,\cite{TVD1} a third phase, that is, by contrast, unstable and hence shall be difficult to be recorded experimentally.

\section{Comparison to experimental results\label{section:fits}}

Interestingly, this phase description in our model matches key observations of fracturing experiments. The abrupt transition, passed a load threshold, from slow cracks to fast cracks, can indeed be interpreted as a phase transition\,\cite{TVD1}, and the usual stick-slip of fronts is a good indicator that some hysteresis holds in the physical laws that rule the rupture dynamics\,\cite{Maugis1985, TVD1}. We then proceed to test our model against two sets of experimental data, where both the energy release rate $G$ and the velocity $V$ of the slow creep and the fast propagation stages were well quantified, as we detail in the next sections.

\subsection{The rupture of PMMA\label{PMMA}}

First, we look into a data set acquired when breaking polymethylmethacrylate plates (PMMA) at room temperature ($T_0 = 296$\,K). A wedge is driven into Perspex\textsuperscript{\textregistered} bodies, resulting in cracks for which two stable ($G$, $V$) branches are indeed recorded\,\cite{PMMA1}. These results are shown in Fig.\,\ref{fig:altuglas}. There, the fast branch, with propagation velocities above $100$ m\,s\textsuperscript{-1}, was reported by \citet{PMMA1}, and the slow creeping branch is here published for the first time for this given PMMA (see appendix\,\ref{method} for details on how it is obtained). When forcing the rupture velocity between these two regimes (i.e., above a specific creep velocity of $4$ cm\,s\textsuperscript{-1} and below $\sim100$ m\,s\textsuperscript{-1}), some stick-slip is observed in the dynamics of the fronts, as reported by \citet{PMA3ss}.\\
Figure \ref{fig:altuglas} then compares both experimental branches with our proposed model. We thus pursue by detailing how each parameter was fitted (i.e., how the model was calibrated to the data), based on asymptotic read-offs. We classically start by wondering how well the slow propagation phase is represented by an Arrhenius law of constant temperature. In the model, this corresponds to a linear $\ln(V)$ to $G$ relationship that holds at low velocity, where $\ln$ is the natural logarithm. There, $\Delta T$ is negligible compared to the background $T_0$ and $G$ is high enough for the healing terms of Eq.\,(\ref{arrh1}) to be secondary (i.e., the terms involving $G_h$ in this equation), leading to
\begin{equation}
\ln (V)=G\left[\frac{d_0^3}{2 \xi k_B T_0}\right] + \left[\ln (V_0) - \frac{d_0^3 G_c}{2 \xi k_B T_0}\right].
\label{linear}
\end{equation}
In the data, this equation shall describe the portion of the plot lying between $10^{-4}$ and $10^{-2}$ m\,s\textsuperscript{-1}, and the slope there, approximately $0.02$\,m\textsuperscript{2}\,J\textsuperscript{-1}, hence constrains $d_0^3/(2 \xi k_B T_0)$ and so the equivalent length for the crack mechanical shielding $\xi$ to be in the order of $50$\,nm. Additionally, the intercept of Eq.\,(\ref{linear}) with the $V$ axis (i.e., the second term in brackets) links $V_0$ and $G_c$. We earlier stated the former to be comparable to the medium Rayleigh velocity\,\cite{PMMA1}, $880$\,m\,s\textsuperscript{-1} in this particular polymer, so that we can deduce the rupture threshold $G_c$ to be about $1300$\,J\,m\textsuperscript{-2}. This value, together with that of $\xi$, gives a fracture energy $U_c=G_c d_0^3/(2\xi)$ comparable to $1$\,eV, which is satisfyingly consistent with a covalence-like barrier. Next, the healing threshold $G_h$ can be inferred from the vertical asymptote at $G=300$\,J\,m\textsuperscript{-2}, below which healing seems to prevail as cracks do not propagate forward\,\cite{RICE1978}. Equation (\ref{arrh1}) predicts this asymptote for $G = (G_c- G_h)/2$, when the healing term equals the breaking one, such that $G_h \sim 650$\,J\,m\textsuperscript{-2}. Let us now focus on the maximum $G$ in the slow stable phase, denoted $G_a$ (for `avalanche') in Fig.\,\ref{fig:altuglas}, around $V=4$\,cm\,s\textsuperscript{-1}. It is modelled by Eq.\,(\ref{arrh1}) once $\Delta T$ is high enough compared to $T_0$ to trigger a phase transition, which, as per Eq.\,(\ref{slow}), mainly depends on the $\lambda/\phi$ ratio. By tuning this ratio, and appreciating the fit (see appendix\,\ref{fit}), we have deduced it to be around $0.9$\,J\,s\textsuperscript{-1}\,m\textsuperscript{-1}\,K\textsuperscript{-1}. As the PMMA conductivity, $\lambda=0.18$\,J\,s\textsuperscript{-1}\,m\textsuperscript{-1}\,K\textsuperscript{-1}, is known\,\cite{altuglas}, we can approximate $\phi \sim 20$\%. Note that at this particular point (at $G=G_a$), the polymer suddenly breaks (e.g.,\,\cite{MaugisBarquins_1978,Parvin1979}), as $\partial V/\partial G \to +\infty$ and the velocity has to jump to the fast regime. Consequently, $G_a$ is often seen as a macroscopic critical energy release rate, which in our description is less than the intrinsic microscopic energy barrier (i.e., $G_a < G_c$). This difference is here directly related to the thermal conductivity $\lambda$ of the medium, and the avalanche to a fast rupture arises when the diffusion can no longer cope with the crack velocity, so that heat is no longer efficiently diffused away from the tip. The characteristic size $l$ on which this heat is generated is the only parameter that remains to be determined. As, according to the model, the crack needs to be hot enough to explain some fast fronts at low mechanical load (i.e., the slower part of the fast branch in Fig.\,\ref{fig:altuglas}, around $100$ m\,s\textsuperscript{-1}), we can estimate the limiting factor of ${\Delta T}_{\text{fast}}$, $Cl$ (see Eq.\,(\ref{fast})). Matching the data set in this area (see appendix \ref{fit}), and using\,\cite{altuglas} $C \sim 1.5\times10^6$\,J\,K\textsuperscript{-1}m\textsuperscript{-3}, we have deduced $l$ to be in the nanometer range. This magnitude happens to be in the same order as the earlier derived $\xi$. We thus predict that most of the induced molecular agitation is introduced on the closest atoms around the crack tip, which coincides with the length scale for the energetic shielding of the tip. Noteworthily, such a nanometer scale appears to be close to the typical entanglement scale of polymers\,\cite{Entangled} (i.e., the density of polymeric chains crossing points in the matrix).\\
To quantify how well the model accounts for the experimental data, we computed, for each data point, the relative orthogonal distance $\varepsilon_d$ to the model, that is
\begin{equation}
\varepsilon_d(G_d,V_d) = \text{min}_m\sqrt{\left[1-\frac{G_m}{G_d}\right]^2+\left[1-\frac{\log_{10}(V_m)}{\log_{10}(V_d)}\right]^2},
\label{error}
\end{equation}
where the subscript $d$ stands for `data' and $m$ for `model'. We are thus looking at a relative fit mismatch along the $G$ axis and a relative fit mismatch, in order of magnitude, along the $V$ axis. For any particular measurement point below $G=G_c$, $\varepsilon_d$ is at most $16$\%. An average error for the whole fit, $\overline{\varepsilon}=\text{mean}_d(\varepsilon_d)$, can also be inferred. To do so, we first
have regularly under-sampled the experimental data onto $40$\,J\,m\textsuperscript{-2} wide bins, keeping there only the mean $G_d$ and the mean $\log_{10}(V_d)$. This way, and doing so separately for the two propagation branches (see appendix\,\ref{meanfit}, Fig.\,\ref{undersampl_altu}), no bias is introduced on $\overline{\varepsilon}$ by the strong difference in measure density along the experimental ($V_d$, $G_d$) curve (i.e., see Fig.\,\ref{fig:altuglas}). The thus derived overall fit error computes to $\overline{\varepsilon}=4$\%, below $G\,=\,G_c$. We discuss, in the next section, the fit beyond $G_c$ and further discuss the accuracy of the inverted parameters in appendix\,\ref{fit}.

\subsection{On the fast crack velocity in PMMA\label{micro}}

\begin{figure}[b]
  \includegraphics[width=1\linewidth]{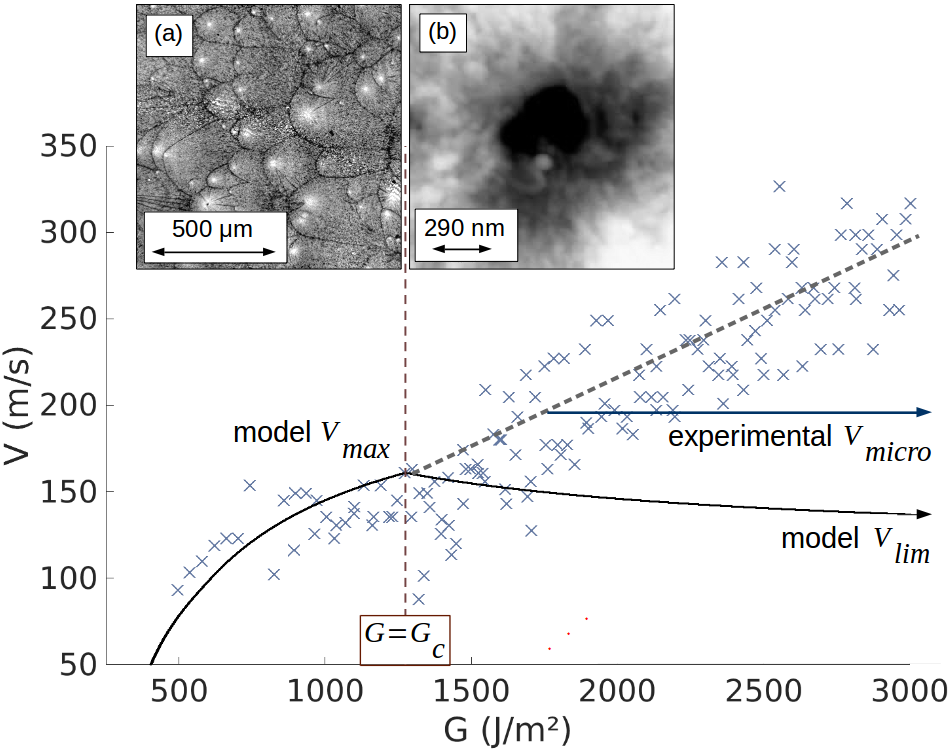}
  \caption{Zoom on the PMMA fast propagation branch presented in Fig.\,\ref{fig:altuglas}, and as per Eqs.\,(\ref{arrh1}) and\,(\ref{tempdiff}). Beyond a load comparable to the modelled $G_c$ threshold, some micro-cracks start to nucleate, impacting the overall propagation velocity as explained by Guerra et al.\,\cite{PMMA2}. The individual velocity of each micro-crack stays however constant at $V=V_\text{micro}$. The validity of our single front model is limited passed this point, although it does predict a velocity plateau $V_\text{lim}$, as per Eq.\,(\ref{vmax}), and a velocity maximum $V_\text{max}$, which are comparable to $V_\text{micro}$. Inset (a): Fractography of the secondary micro-cracks on a postmortem fracture surface. White areas mark their nucleation centres. Inset (b): Atomic Force Microscopy of a nucleating cavity at the centre of a micro-crack. As proposed in section \ref{concl3}, it could derive from the sublimation of localised bubbles around the main front, due to some intense thermal effects.}
  \label{fig:micro}
\end{figure}
Our simple sub-critical model hence matches most of the rupture dynamics of PMMA, from slow to fast velocities. In Fig.\,\ref{fig:altuglas} however, an increase in velocity holds in the experimental data beyond $G=G_c$, and is not properly accounted for. To highlight this mismatch, we display in Fig.\,\ref{fig:micro} the fast branch with an optimised display scale.
It has been shown\,\cite{PMMA2} that, beyond a particular load, the global front velocity is impacted by the fracture instabilities that occur at high speed. Indeed, passed this threshold, fronts get more complex as micro-cracking occurs\,\cite{Smekal1953,RaviChandar1997,PMMA1}, that is, as micro-cracks form and propagate in the fracture plane ahead of the main front. Such micro-cracks are shown in Fig.\,\ref{fig:micro}. And, at an even higher load, micro-branching also comes into play, and aborted out-of-plane secondary cracks are observed\,\cite{RaviChandar1984,Fineberg1991,Sharon1995,Boue2015}. For the PMMA that is here studied, the micro-cracks were observed\,\cite{PMMA1} at velocities above $165$\,m\,s\textsuperscript{-1}, which approximately corresponds in the model to $G>G_c$. Beyond this threshold, the apparent macroscopic speed of the front, $V$, increases with the micro-cracks growing density, while the individual velocity of each micro-front, however, was inferred to stay constant\,\cite{PMMA2}, around $V_\text{micro}\,\sim 200\,\mathrm{m\,s}^{-1}$, as illustrated in Fig.\,\ref{fig:micro}.
\\Such a plateau in the propagation speed is somewhat consistent with our description (see Fig.\,\ref{fig:micro}). But this being said, it is clear that our unique front model shows limitations as soon as fronts complexify.
We can still push this discussion on the fast regime a bit further. A question of interest about the rupture of PMMA has been why the maximal observed crack velocity was significantly lower than the theoretical Rayleigh speed\,\cite{PMMA1} (i.e., about $200$\,m\,s\textsuperscript{-1} rather than $880$ m\,s\textsuperscript{-1}). Equation (\ref{arrh1}) gives here some insight, as it does predict a plateau velocity $V_\text{lim}$ as the applied $G$ gets very large. Indeed, besides preventing the crack advance at very low loads, the sub-critical healing processes significantly limit the fast growth rate, as the tip temperature is modelled to be high. More specifically, by inserting ${\Delta T}_{\text{fast}}$ (\ref{fast}) in Eq.\,(\ref{arrh1}), and by looking at the high loads asymptotic regime of the healing term, we predict $V$ to be limited by $c_0=d_0^3 C$, the individual heat capacity of atom bounds:
\begin{equation}
\begin{split}
   V_\text{lim} &\sim V_0 \Bigg[ 1 - \exp \left( {-\frac{d_0^3(1+\bcancel{G_h/G})}{2\xi k_B (\bcancel{T_0/G}+\phi/[\pi C l])}} \right) \Bigg]\\\\
   &\sim V_0 \Bigg[ 1 - \exp \left( {-\frac{\pi c_0}{2 k_B}\hspace{2mm}\frac{l}{\xi\phi}} \right) \Bigg].
   \label{vmax}
\end{split}
\end{equation}
In this expression, the crossed out terms are neglected in regard to the neighbouring ones. 
Note however that Eq.\,(\ref{vmax}) is mainly illustrative, as the plateau it describes occurs in a domain where our single front model does not strictly apply. Note also that the value $V_\text{lim} \sim 100$\,m\,s\textsuperscript{-1} is smaller than the modelled maximum individual propagation velocity $V_\text{max} \sim 160$\,m\,s\textsuperscript{-1}, which is obtained for $G=G_c$ rather than for $G \to +\infty$ (see Fig.\,\ref{fig:micro}).

\subsection{The detachment of Pressure Sensitive Adhesives}

\begin{figure*}
  \includegraphics[width=1\textwidth]{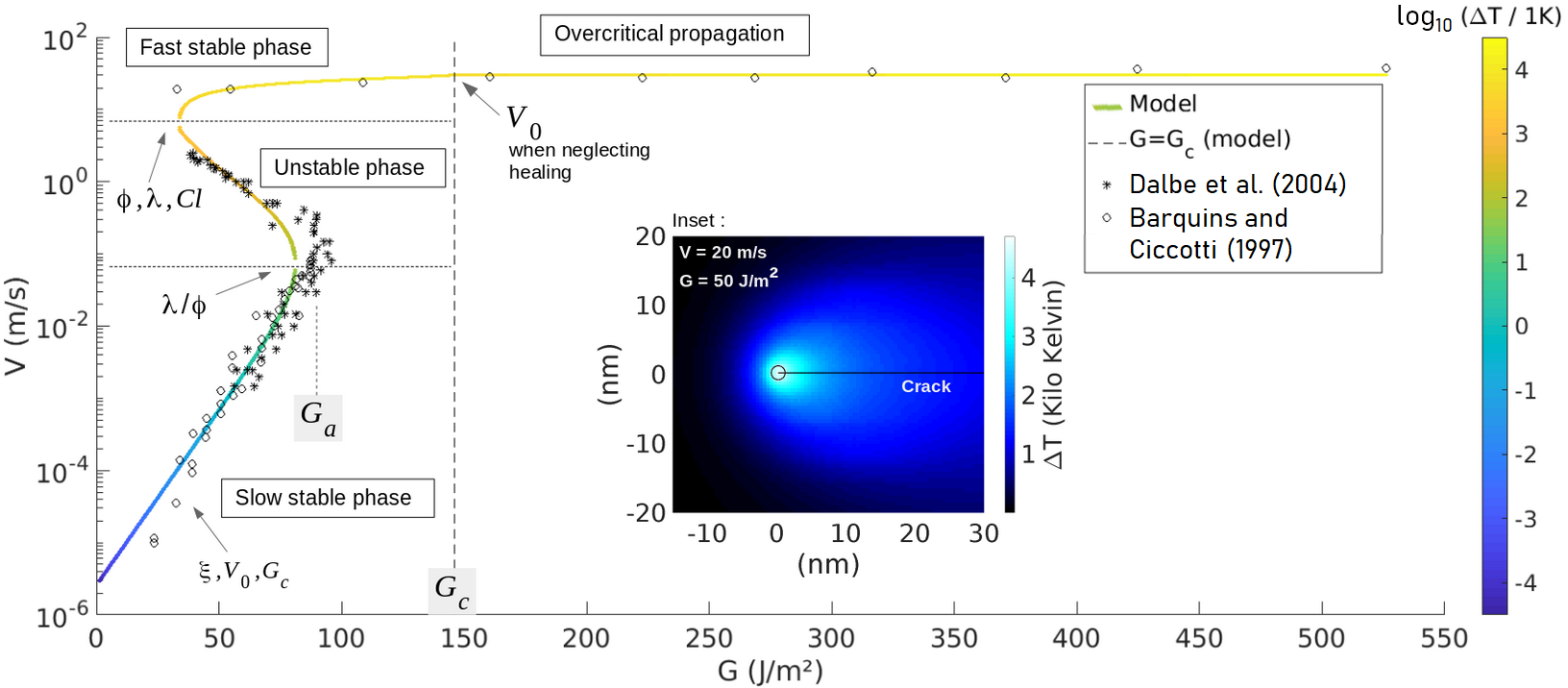}
  \caption{Crack velocity $V$ as a function of the energy release rate $G$ as predicted by Eq.\,(\ref{arrh1}) and (\ref{tempdiff}) (plain curve) and fitted to the tape experimental data\,\cite{tape1,tape2}. The unstable branch was not actually measured and the data points there are only averaged $V$ versus $G$ for a crack that undergoes stick-slip, in the given set-up, between the slow and the fast phase. The arrows indicate to which model parameters each part of the curve is mainly sensitive, and the main color scale specifies at which temperature the crack tip is modelled to be. The load $G_a$ is an avalanche threshold beyond which peeling fronts can only propagate quickly and $G_c$ is the modelled microscopic energy barrier for rupture. The inset shows the associated modelled temperature field around the front, at the onset of the fast to slow phase shift ($G=50$\,J\,m\textsuperscript{-2}, $V = 20$\,m\,s\textsuperscript{-1}). For readability, the color map is there different from the main one, and the circle corresponds to the tip of radius $l$, where the extra heat is emitted. $\Delta T$ of the main model curve is the value at the centre of the circle.}
  \label{fig:tape}
\end{figure*}
We now pursue the comparison with the reported rupture of another material, acrylic based pressure sensitive adhesives (PSA), that typically happens when unrolling some office tape. In particular, the peeling dynamics of Scotch\textsuperscript{\textregistered} 3M 600 rolls (composed of a polyolefin rigid backing coated with a layer of synthetic acrylic adhesive) has been thoroughly studied in the last decades (e.g.,\,\cite{tape1, tape2, PRL_tape}); we here fit our model to two compatible ($G$, $V$) data sets that were published by \citet{tape1} and by \citet{tape2}. These data sets are shown in Fig.\,\ref{fig:tape}. Two stable modes of front detachment (i.e., a fast one and a slow one) are reported\,\cite{tape2}, similarly to those governing the rupture in PMMA. Additionally, some (unstable) stick-slip in the rupture dynamics is also observed\,\cite{tape1} when peeling with an average velocity between $V \sim 15$\,cm\,s\textsuperscript{-1} and $V\sim20$\,m\,s\textsuperscript{-1}.\\
Overlaying this experimental data, Fig.\,\ref{fig:tape} also displays a calibrated version of our model. The model parameters were inverted as follows, with a similar asymptotic analysis as what was done for PMMA. As no significant healing threshold displays at low velocity, we have only assumed that $G_h$ is high enough to completely neglect the healing processes (i.e., the healing term in Eq.\,(\ref{arrh1}) is small if $G_h$ is high). Of course, this absence of threshold, below which no forward propagation of the crack is observed, could also only indicate that $(G_c-G_h)/2<0$ or that this value (i.e., illustrated on the PMMA data in Fig.\,\ref{fig:altuglas}) is less than the minimum energy release rate that was investigated in the tape experiments. We discuss this particular point further in appendix \ref{ap_tape}. We now invert the length $\xi$, which is, again, given by the slope of the slow phase and is here about $10$\,nm. As no healing is now supposed to be at play, the nominal velocity $V_0$ is given by the highest velocity records: $V_0 \sim 30$\,m\,s\textsuperscript{-1} as $V_0$ is the maximum value then predicted by Eq.\,(\ref{arrh1}). Satisfyingly, this value compares well with the magnitude of a mechanical wave velocity in PSA, that is, $\sqrt{\mu/\rho}$, where $\mu$ is, for instance, the shear modulus of the adhesive\,\cite{PSA}, $0.1$ to $1$\,MPa, and $\rho$ is its volumetric mass\,\cite{PSA2}, about $10^3$\,kg\,m\textsuperscript{-3}. Next, from Eq.\,(\ref{linear}), the intercept of the slow branch with the ordinate (zero $G$) axis indicates $G_c \sim 150$\,J\,m\textsuperscript{-2}. Rather logically, and with the inverted value of $\xi$, this again corresponds to a value of fracture energy $U_c\sim1$\,eV. Note also that $G_c$ is again higher than the transition load $G_a$ at which a creeping front jumps to a fast regime. From this transition load, arising in the model from the temperature rise at low velocity\,(\ref{slow}), we also infer $\lambda/\phi$ to be in the order of $0.1$\,J\,s\textsuperscript{-1}\,m\textsuperscript{-1}\,K\textsuperscript{-1}. As the adhesive's conductivity $\lambda$ lies in the same range\,\cite{PSAtherm}, a consequent portion of $G$ should be released into heat: $\phi \sim 1$. Of course, $\phi$ cannot be exactly one, as other dissipating processes than heat diffusion are likely to dissipate a part of $G$ (see the discussion in section \ref{concl}). According to our inversion however, this part ought to be small. Finally, by varying $l$ and by matching the coolest points of the fast phase, we estimate this parameter, which limits the highest tip temperature (i.e., Eq.\,(\ref{fast})), to be in the nanometer range. This value, for the length scale of the heat production zone, is again rather consistent with the inverted magnitude of $\xi$, that is, the equivalent length scale for the mechanical shielding of the tip. Note also that both $\xi$ and $l$ are interestingly comparable to what was obtained for PMMA, and in the order of a polymeric entanglement density\,\cite{Entangled}.\\
As shown in Fig.\,\ref{fig:tape} and with this set of parameters, the model accounts for most of the tape peeling dynamics. More quantitatively, for all the particular data points of the two stable phases, the fit error $\varepsilon_d$ (as defined by Eq.\,(\ref{error})) is less than $20$\%. We also computed a mean fit error $\overline{\varepsilon}=\text{mean}_d(\varepsilon_d)$ for the stable phases. To do so, and as done for PMMA, we first averaged the data points onto $10$\,J\,m\textsuperscript{-2} wide bins, so that no densely populated part of the measured curve dominate the value of $\overline{\varepsilon}$ (see appendix\,\ref{meanfit}, Fig.\,\ref{undersample_tape}). We thus computed $\overline{\varepsilon}=5$\%.\\
Note that, in comparison to the fast branch for the failure of PMMA (i.e., as discussed in section\,\ref{micro}), it would be of interest to know if the critical load $G=G_c$ also approximately corresponds to the apparition of some new rupture modes. Yet, the high velocity branch of the tape data is bound to relatively large uncertainties (the loading system of \citet{tape2} involved dropping weights from an elevated balcony, illustrating the challenges in fast peeling measurements), so that it does not allow a more thorough analysis.

\newpage
\subsection{Parameter summary}

In Tab.\,\ref{table}, we summarises all the parameters's values, that we have inverted or supposed for the rupture of PMMA and PSA. The accuracy of these values is further discussed in appendix\,\ref{fit}.
\begin{table}[ht]
\begin{tabular}{c|c c|c}
\hline
\textbf{\,Parameter\,} & \textbf{\,\,\,\,\,\,\,\,\,PMMA\,\,\,\,\,\,\,\,\,} & \textbf{\,\,\,\,\,\,\,\,\,PSA\,\,\,\,\,\,\,\,\,} & \textbf{\,\,\,\,\,\,\,\,\,Unit\,\,\,\,\,\,\,\,\,} \\ \hline \hline
$V_0$               & $880$           & $30$  & m s\textsuperscript{-1}        \\ \hline
$G_c$               & $1300$          & $150$ & J m\textsuperscript{-2}         \\ \hline
$G_h$               & $650$          & - & J m\textsuperscript{-2}         \\ \hline
$\xi$                 & $50$           & $10$ & nm           \\ \hline
$l$                  & $1$             & $1$ & nm            \\ \hline
$\phi$                & $0.2$          & $\sim1$ & [ - ]           \\ \hline
$\lambda$             & $0.1$          & $0.18$ & \,J s\textsuperscript{-1} m\textsuperscript{-1} K\textsuperscript{-1}\,        \\ \hline
$C$                  & $1.5$        & $1$ & MJ m\textsuperscript{-3} K\textsuperscript{-1}            \\ \hline \hline
$G_c/G_a$               & $1.8$          & $1.6$ &   [ - ]     \\ \hline
$U_c$               & $1$          & $1$ &   eV     \\ \hline
$N$               & $500$          & $100$ &   [ - ]     \\ \hline
\end{tabular}
\caption{Summary of all model parameters considered for the rupture of PMMA and PSA, as discussed in section\,\ref{section:fits}. A value $d_0 \sim 2$\,{\AA} has been assumed in the derivation of these parameters. For completeness, the related quantities $G_c/G_a=U_c/U_a$, $U_c=G_c d_0^3/(2\xi)$ and the shielding factor $N=2\xi/d_0$ are also specified.}
\label{table}
\end{table}

\section{Discussions\label{concl}}

For two different polymeric materials, we thus have shown how a thermally activated fracture process, coupled with the dissipation and diffusion of heat, can simply explain many features of the dynamics of both creeping and fast cracks, and the shifts from one state to the other. Such novel match, over seven to nine decades of propagation velocities and with only very simple physics considerations, could shade some new light on fracture mechanics, as thermal effects are often discarded.

\subsection{How hot is too hot for a crack tip? Some light from fractoluminescence}

\begin{figure}[b]
  \includegraphics[width=1\linewidth]{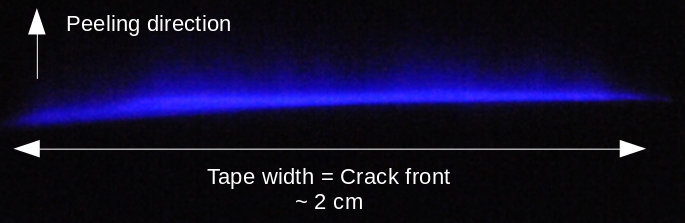}
  \caption{Blue radiation emitted when quickly peeling tape beyond the stick-slip threshold (i.e., at an average velocity greater than $15$ cm\,s\textsuperscript{-1}, see Fig.\,\ref{fig:tape}). This picture was captured in the dark by a standard reflex camera (ISO: $25600$, shutter speed: $1/2$ s, focal length: $60$ mm, aperture: f/4). The low shutter speed ensures that enough light enters the camera, but then covers many stick-slip cycles of the peeling dynamics\,\cite{tape1}. Such fractoluminescence could be the mark of a very hot crack front\,\cite{WEICHERT_1978, tribo_blackbody, Bouchaud2012} when unrolling tape.}
  \label{fig:flash}
\end{figure}

To explain the fast propagation branch, we have notably predicted the front temperature to reach several thousands of degrees. Such high values are difficult to confirm experimentally, especially as they are to stand only on a few nanometers during short avalanches. There exist however, indirect hints toward the existence of an important temperature elevation in a variety of brittle materials fracturing at high speed.\\
For instance, the analysis of some fracture roughness in cleaved quasi-crystals has revealed a damage zone of size anomalously large for this class of materials, and this was stated to result from a local temperature elevation of about $500$\,K at the moving crack tip\,\cite{BonamyTemp}.\\
Several experimental works in glass and quartz\,\cite{WEICHERT_1978, tribo_blackbody, Bouchaud2012} also managed to indirectly measure $\Delta T$ to indeed reach thousands of degrees, by characterising the photons emission from the tips of some moving cracks and by comparing it to the blackbody radiation theory\,\cite{blackbody}. In the case of tape, when peeling fast enough to be in the stick-slip regime, a blue tribo-radiation can similarly be observed\,\cite{tape2,natX}, and it was established that this radiation only occurs during the fast propagation phases of the cycle\,\cite{tape2}. A direct example of such an emission is shown in Fig.\,\ref{fig:flash}, and its color could well correspond to the central wavelength $\lambda_\text{peak}$ associated, via Wien's law\,\cite{blackbody}, with a blackbody temperature compatible with our model:
\begin{equation}
   \lambda_\text{peak}=\frac{b}{T_0+\Delta T}\sim400\mathrm{\,nm},
   \label{wien}
\end{equation}
where $b$ is Wien's displacement constant $\sim0.0029$\,m\,K and $\Delta T$ is about $7000$\,K at a load just passed the stick-slip threshold $G=90$\,J\,m\textsuperscript{-2} (see Fig.\,\ref{fig:tape}).
The intensity of the observed light, which is visible in the dark but not under normal lightening, seems to also be consistent with the model. According to the Stefan–Boltzmann law\,\cite{blackbody}, we indeed expect a radiated power in the order of
\begin{equation}
P = s(T_0+\Delta T)^4hl \sim 1\,\mathrm{mW},
   \label{stefan}
\end{equation}
where $s\sim5.67 \times 10^{-8}$\,W\,m\textsuperscript{-2} K\textsuperscript{-4}
is the Stefan–Boltzmann constant, and $h$ the tape width ($2$\,cm) so that $hl$ is the total area that significantly emits light. Note that such a power only accounts for a negligible part of the energy that is dissipated as the front advances, as $P/(GVh) \sim 10^{-4}$. For a human eye at a distance $D \sim 10$\,cm, it corresponds to a light luminance of about $eP/(4\pi D^2) \sim 1$\,cd\,m\textsuperscript{-2}, using a blackbody luminous efficacy\,\cite{blackbody_efficacy} $e$ of $100$\,lumens per watt. With a pupil opening of about $10$\,mm\textsuperscript{2}, such a luminance is in the order of $10$\,trolands (Td)\,\cite{inthedarklight}, which does fit that of a flickering (i.e., the front has a stick-slip motion) radiation that is only visible in the dark, as those approximately range between $0.01$ and $100$\,Td\,\cite{inthedarklight}. While the eye is persistent, a camera sensor of size $S \sim 10$\,mm\textsuperscript{2}, placed at the same distance, would capture an averaged power $\gamma PS/(4\pi D^2) \sim 100$\,nW, where $\gamma \sim 0.1$ is a typical ratio of time during which the front is in the fast phase compared to the total recording time, when peeling at a slow average velocity (i.e., $\sim15$\,cm\,s\textsuperscript{-1})\,\cite{tape1}. The magnitude of this power is interestingly close to the $10$\,nW that were successfully measured by \citet{natX} for the luminescence of another adhesive roll.
\\For a given PMMA, \citet{Fuller1975} also tried to quantify the temperature elevation around a quick fracture, both with the thermoluminescence technique and by using a liquid crystal coating on the matrix, whose color was thermosensitive\,\cite{thermo_crystal}. For cracks propagating at $400$\,m\,s\textsuperscript{-1} and faster, they measured heat efficiencies of about $2000$ J\,m\textsuperscript{-2}, which is fairly compatible with the value we have derived for $\phi G$ (a $400$\,m\,s\textsuperscript{-1} speed is obtained for $G>4000$\,J\,m\textsuperscript{-2} in Fig.\,\ref{fig:altuglas} and $\phi$ was inferred to be about $0.2$). This experimental work\,\cite{Fuller1975} also estimated the instantaneous temperature elevation of the fractures to be about $500$\,K over a $0.5$-$\upmu$m-thick area around the front. Such a thickness for the heat source was however acknowledged to be rather uncertain, as the measure sensibility for this parameter was limited. We remark that the same energy spread on the $l\sim10$\,nm thickness which we have here inferred would give a temperature rise of $10^4$\,K and more, as predicted by our model (see Fig.\,\ref{fig:altuglas}).
\\Truly, fractoluminescence could emanate from other mechanisms than some hot matter radiation. It was for instance proposed\,\cite{natX, electric_tape} that it partly arises from the molecules excitation of the fracture in situ air, by some electrical discharges between the two crack planes. Both these phenomena could surely coincide and, in any case, the light emission is an indication that some extreme and localised phenomena are at stake during fast failure. In that way, the thermodynamics model we propose holds some compatibility with that of \citet{Slepyan_81}, where the abrupt advance of cracks derives from the emission of high frequency phonons, that excite atom bonds ahead of the tip, but that do not necessary thermalize.\\
Some relatively recent atomistic simulations\,\cite{atomheat} seem nonetheless to confirm that the atoms at a moving front can undergo a significant heat. In a modelled graphene, \citet{atomheat} thus inferred a $200$\,K temperature rise, over a $43$\,nm\,$\times\,43$\,nm area surrounding a running tip. This estimation is interestingly compatible with the thermal maps presented in Figs \ref{fig:altuglas} and \ref{fig:tape}, for which the mean temperature is respectively $950$\,K and $350$\,K, when recomputed on a similar $1800$\,nm\textsuperscript{2} surface upon the front. Note that atomistic simulations might naturally be more proper than our mesoscopic description, in particular because the small scales ($l$) and high excitation frequencies ($V/l$) at play could call for more complicated models\,\cite{non_fourier, ThermoluminescenceSolids} than plain Fourier diffusion, Arrhenius growth or blackbody radiation. Yet, atomistic simulations are by nature far heavier to run, requiring an accurate description of the atomic interactions onto femtosecond time steps.

\subsection{Is a simple model too simple?\label{concl1}}

It is actually surprising that the proposed simple mesoscopic model can describe the propagation of cracks, when such a propagation, in reality, displays many complex phenomena. For instance, we have completely neglected the impact of crazing on the crack dynamics\,\cite{Crazing, fibril}, that is, the formation of defaults and fibrils at relatively large scales around the fracture front (i.e., a hundred of micrometers in PMMA and up to millimeters in PSA), while such large scale plasticity is often considered to have a strong effect on the growth of cracks (e.g.,\,\cite{Crazing, fibril}). Yet, crazing is not incompatible with our thermal weakening model, which only states that a significant part of the mechanical energy should be dissipated far closer to the crack front (i.e., over a few nanometers), and that this very local dissipation should be that of a first effect on the crack dynamics. In this description, crazing is then a consequence of the front progression rather than its main cause. In a similar way, many other known failure phenomena, such as the emission of mechanical waves during rupture\,\cite{crackwaves}, complicated creep laws from the corrosive interactions between the fracture fluid and the fracture tip (e.g.,\,\cite{lawn_1993}, chpt. 5.4), or the complexification of fronts at high propagation velocities\,\cite{RaviChandar1997}, are not directly encompassed by Eqs.\,(\ref{arrh1}) and (\ref{tempdiff}), but are not in conflict with the model either.\\
The simplicity of the model can actually be considered as one of its strength, as the physics that it describes could apply to many different materials and not only to polymers. Accurately testing this idea would however require the full ($G$, $V$) curves of more materials, and those are often not trivial to obtain experimentally at all velocities. Such experimental work could yet be rewarding, as we have here shown that matching the model to some ($G$, $V$) curves can give some valuable insights on the rupture of matter.
Our quantification for each model parameter stays however rather approximate, and we have mainly derived their orders of magnitude. We have, in particular, assumed that they were all constant for a given material, while most could be velocity or temperature dependent\,\cite{Marshall_1974, carbonePersson, Kitamura2008}. For instance, the fact that PSA exhibits a larger scale viscous behavior (i.e., including fibrillation and heating over millimeters around the tip) at lower velocity\,\cite{fibril} could indicate that the heat production size $l$ decreases with the crack speed in this medium. It is especially known that the elastic moduli in PSA are strongly temperature dependent\,\cite{PSA}, and this was actually proposed by Maugis\,\cite{Maugis1985} and \citet{carbonePersson} as the driving cause for failure instability in rubber-like materials. We have, besides, considered both PMMA and PSA as homogeneously tough while $G_c$ is bound to present some quenched disorder. While such heterogeneities should not affect the stable propagation branches, as long as $G$ and $V$ are then understood quantities which are averaged over a few $G_c$ correlation lengths, it could be of importance for the accuracy of the loads at which the phase transitions occur\,\cite{TVD1}, as slow cracks shall preferentially avalanche on weaker zones and fast cracks stop on stronger locations. In the case of PSA, we have furthermore considered that peeling was a cohesive process (i.e., that it occurs inside the adhesive), while a bi-materials interfacial model would be more appropriate, as the crack essentially propagates at the interface between the substrate and the glue\,\cite{cohesion_interface}.\\
These numerous limitations being stated, the parameters we have inverted are nonetheless in rather satisfying orders of magnitude, confirming the physical relevance of the model. Indeed, the intrinsic fracture energy in both materials $U_c=d_0^3 G_c / (2 \xi)$ is comparable to one electronvolt, which is typical for an energy that bonds atoms (e.g., see appx. E in Ref.\,\cite{covalence}). Because our proposed description is statistical, one should remember that 
$U_c$ is a mean material feature, for a rupture process that is made of several types of bond breaking. As a rough example, $U_c\sim1$\,eV may indicate that the crack consummates in average three weak links (such as hydrogen or Van des Waals bonds of respective energies\,\cite{covalence} $\sim0.1$ and $\sim0.01$\,eV) for every stronger connection that snaps (say, one C--C link of an acrylic chain, of covalence energy\,\cite{covalence} $\sim4$\,eV). The nanometric scale $l$ for the heat generation may well correspond to the typical entanglement density in polymers\,\cite{Entangled} (the density of polymeric chains crossing points in the matrix), below which atoms have more freedom to vibrate, and which is known to affect some rupture properties (e.g.,\,\cite{Entangled, Crazing}). It is also coherent that the generation of heat was inferred to occur over a length scale comparable to $\xi$, the equivalent radius describing the energy shielding of the tip. We have indeed derived that the former is a strong cause for the latter, as the heat efficiency $\phi$ was inverted to be non negligible (i.e., $\phi\sim0.2-1$).

\subsection{Tip stress and front shielding\label{concl3a}}

A nanometric scale (i.e., comparable to $\xi$ or $l$) has been noteworthily observed in the rupture of other materials. One example is the length scale of a light radiating (and hence likely thermal) zone around running fracture tips in glass\,\cite{Bouchaud2012}. In carbonate rocks, it is also the typical size of some observed nanograins that form along sliding seismic fault planes\,\cite{nanoparticles}. Such a nano-damage explains the glossy and reflective aspect displayed by some faults (often referred to as fault mirrors), as their typical surface roughness is then comparable to the wavelengths of visible light. The origin of this damage, however, is debated as, below $1\,\upmu$m, plasticity is expected to dominate over brittleness in this material and asperities should hence deform rather than break. Noteworthily, some intense thermal effects, arising from the frictional heat, such as some fast melting and cooling or the thermal decomposition of carbonates, were proposed to solve this apparent paradox\,\cite{nanoparticles_decompo}.\\
Similarly, for the materials that we have here studied, the usual predictions for the size of the shielding process zones are far larger than $\xi$. In PMMA, for instance, it is in the order of $\xi_\text{macro}\,\sim\,GE/\sigma_y^2 \sim 200$\,$\upmu$m, where $\sigma_y \sim 100$\,MPa is the tensile yield stress of the bulk polymer and $E \sim 3$\,GPa its Young modulus\,\cite{altuglas}. However, in that description, $\sigma_y$ is a stress that is averaged over a macroscopic sample, and is likely not representative of the actual energy density around the defaults of this sample. It was notably reported that a Dugdale\,\cite{dugdale1960} like cohesion model (i.e., $\sigma$ is homogeneously equal to $\sigma_y$ in a process zone of radius $\xi_\text{macro}$), poorly accounts for fast rupture in PMMA\,\cite{PMMA_heterogeneousZone}. Naturally, $\xi_\text{macro}$ is still to bear some significance, in particular as a characteristic length scale for crazing in acrylic glass\,\cite{Crazing}, where a portion within $(1-\phi)$ of the release rate $G$ is to be dissipated, either by the creation of dislocations\,\cite{rice_surface}, the emission of waves\,\cite{Bouchaud2012,crackwaves} or residual thermal effects. But $\xi$ was inverted as an equivalent size, only defined by $2\xi/d_0=N$ with $N$ the damping of the tip potential energy $U$ due to the energy dissipation. We solely inverted $N$ to be around $100$ and $500$ for respectively PSA and PMMA, and many links might well snap and heal far away from the tip, allowing for crazing.\\
Still, most of the rupture is likely to occur very close to the front where the stress is to be the highest. We can estimate such a stress at the tip by considering a simplified expression for the elastic energy stored in rupturing bonds:
\begin{equation}
   U\sim d_0^3\frac{\sigma^2}{2E},
   \label{U}
\end{equation}
which, with Eq.\,(\ref{UG2}), is equivalent to the well known form for the limitation of an otherwise divergent stress at the tip of cracks, predicted by the general elasticity theory (e.g.,\,\cite{lawn_1993}):
\begin{equation}
    \sigma \sim \sqrt{\frac{GE}{\xi}}.
    \label{sigG}
\end{equation}
In the case of PMMA, such a computed stress is as high as $7\,$GPa, and we thus predict a high atomic strain $\sigma/E$ of about $200$\% at the onset to fast rupture (i.e., for $U$ equal to $U_c/1.8$ as per Tab.\,\ref{table}). Such a strain shall be likely at a fracture tip for the strong intermolecular deformation immediately before failure. Of course, the simply linear elastic Eq.\,(\ref{U}) is unlikely to be valid at $200$\% strain, and we also considered describing $U$ with a Morse potential\,\cite{Morse}, that is
\begin{equation}
    \frac{U}{U_c}\sim\left(1-\exp\left[-\sqrt{\frac{Ed_0}{2U_c}}(d-d_0)\right]\right)^2,
    \label{Morse}
\end{equation}
which, at the onset of fast rupture, predicts a strain $(d-d_0)/d_0\sim400$\%. While this dual-particles potential stays, by nature, a strong approximation in the complex rupture of a polymer, it is worth reminding that the model which we have introduced does not rely on a particular shape of the inter-atomic potentials (i.e., neither on Eq.\,(\ref{U}) or on Eq.\,(\ref{Morse})), but only on their average dissociation energy\,$U_c$.

\subsection{Front complexification\label{concl3}}

Overall, our derivation of $\xi \ll \xi_\text{macro}$ only suggests that process zones are heterogeneous objects, dissipating a higher density of energy  in their centre than at their periphery. In particular, it was shown that a few tens of micrometers (i.e., a portion of $\xi_\text{macro}$) is a typical distance at which the secondary micro-cracks nucleate from the main front in PMMA\,\cite{PMMA2} and, as shown in Fig.\,\ref{fig:micro}, the imaging of some postmortem rupture surfaces reveals that these micro-cracks initially grow from isolated spherical cavities at their centre, of radius about $300$\,nm. We here propose that such cavities could correspond to bubbles, forming by sublimation\,\cite{bubble_kinetics} on weak locations of the process zone, and leading to some micro-fractures once having grown to a critical size. While remaining to be confirmed, such a sublimation process would definitely require some local but very high temperatures in the crazing area. Indeed, to nucleate ahead of the main front, the observed cavities have to form during less than $\xi_\text{macro}/V\sim1$\,$\upmu$s, and the pyrolysis of PMMA to methyl methacrylate (MMA) only reaches such a reaction rate at temperatures $T_b$ that are beyond $1000$\,kelvins\,\cite{bubble_kinetics}. In return, and assuming that the ideal gas law approximately applies (e.g.,\,\cite{kinetics}, chpt. 4), some bubbles forming at this temperature would hold an internal pressure $ \rho R T_b / M$, where $M=0.1$\,kg\,mol\textsuperscript{-1} is the MMA molecular mass\,\cite{periodic_table}, $\rho=1200$\,kg\,m\textsuperscript{-3} is the volumetric mass of the solid PMMA\,\cite{altuglas} and $R$ is the ideal gas constant. This value computes to at least $100$\,MPa, which is comparable to the surrounding bulk compressive strength\,\cite{altuglas}. The evolution from pressurised pores to propagating micro-cracks would then be coherent.
\\Thus, in addition to explaining, as shown in this work, the first order dynamics of singular fronts, concentrated thermal processes could also be responsible for their complexification at high propagation velocities. In the case of acrylic glass, we have notably inferred (see section\,\ref{micro}) that the appearance of the secondary fronts approximately coincides with energy release rates that are close to the (modelled) intrinsic barrier $G_c$. This concomitance could be explained by the need for new dissipation processes, when cracks propagate over-critically ($G>G_c$) so that some extra energy is brought to the rupture system. Such an idea is notably re-enforced by the fact that the density of nucleated micro-cracks was inferred to be proportional to a value comparable to $G-G_c$, as shown by \citet{PMMA2}.

\section{Conclusion and perspectives}

We presented a new and general model for the kinetics of cracks. The main physical elements that were introduced in this model are, only, a sub-critical (Arrhenius-like) growth rate and the dissipation and diffusion of heat around fracture tips (where the applied mechanical stress is concentrated), an immediate consequence of the latter being the possibility for crack fronts to reach thousands of degrees temperatures. Interestingly, these different elements have, separately, long been considered or observed in the physics of rupture (e.g.,\,\cite{Brenner, Marshall_1974, BARENBLATT_1962, ricelevy1969}), but had not previously been combined for comparison with some experimental data. In doing so, we here showed that the rupture of two materials (namely, PMMA and PSA) can be quantitatively reproduced over many decades of propagation velocities, from slow creep regime to fast propagation.\\
Thus, we inferred that the propagation of a crack can be sub-critical, even at velocities approaching that of the mechanical waves in the surrounding matrix, due to its potentially very high tip temperature. We also suggested that the microscopic healing process around a fracture front can significantly constrain the fast velocity regime, from the strong thermal activation at such temperatures, while it is often considered that healing is only relevant for very slow cracks. The existence of thousands of degree temperatures is actually supported by many experimental works that study the visible fractoluminescence of fast fronts\,\cite{WEICHERT_1978, tribo_blackbody, Bouchaud2012,Fuller1975}. In some instance\,\cite{atomheat}, it has also been modelled by some atomistic simulations, and we additionally showed, in the present work, the existence of bubble forming in the process zones of cracks in PMMA. We proposed that these bubbles could well originate from some local sublimation of the polymer near crack tips. As they are located at the nucleation centres of secondary fracture fronts, we also suggested that the complexification of cracks at high velocities could derive, as the rest of the propagation dynamics, from some thermally activated processes. Finally, for the two materials that we have studied, we have inferred that the mechanical stress around cracks remains an increasing quantity inside the process zones up to a few nanometers from the tip. Such a nanometric scale matches the typical size over which the heat was inferred to be generated, making thermal dissipation the likely main process that shields rupture fronts from mechanical failure.\\
In theory, the model could be reversed, and the fast propagation of cracks under a minimum load could be triggered by a very local heating of their fronts. Related experiments could for instance be performed with localized light pulses on material that are thin or transparent enough, for the heat elevation to be controlled. Certainly, elevated ambient temperatures are known to have a strong impact on the kinetics of fractures, both in the lab (e.g.,\,\cite{creepBaud,temp_PMMA}) and in nature\,\cite{cliffHeat}.\\
Noteworthily, the proposed model, and its ability to explain some actual crack dynamics, stresses the importance of the heat conductivity of materials on their macroscopic strength. A high conductivity indeed allows to evacuate the extra internal energy away from the fronts, thus delaying any thermal weakening. As a general statement, many strong materials happen to be good conductors, such as metals, graphene\,\cite{Graphene_cond} or spider silk\,\cite{Spider_Cond}. For the latter, it was in particular shown that, contrarily to most materials, its conductivity actually increases with deformation\,\cite{Spider_Cond}, which could well be a natural defence mechanism for the stability of arachnid webs. Designing human-made solid matrices that can replicate such a behavior on the molecular scale could then become a new important target of material sciences.\\
Finally, we suggest that most of the physics that we have introduced to study mode I fractures shall also be valid for mixed-mode fracturing as well as for solid friction. The latter is actually suspected to hold some non negligible, thermal related, weakening mechanisms (e.g.,\,\cite{HeatWeak}), which could notably be a key in geophysics in understanding the stability of seismic faults. Such mechanisms might be diverse, and may include the thermal pressurisation of fault fluids\,\cite{ratestate, pressur2005} or some changes in the fault planes minerals phase (i.e., such as melting or thermal decomposition)\,\cite{SulemCarbo}. We propose that they could also be related to a thermally boosted sub-critical slip, in the sense of statistical physics and similarly to the model we have here developed.


\newpage
\section*{Acknowledgements and contributions}
\noindent T.VD. developed and analysed the model and redacted the manuscript. R.T. proposed the physical basis of the model and its mathematical formulation. A.C. worked on its numerical implementations. D.B. and L.H. provided the PMMA data and images. S.S. and L.V. gave direct insights on the tape experiments. K.J.M. and E.G.F. contributed in the interpretation of the model in fracture mechanics applications. All authors participated to the redaction of the manuscript and agreed with the submitted version. The authors declare no competing interests in the publishing of this work. We acknowledge and are grateful for the support of the IRP France-Norway D-FFRACT, of the Universities of Strasbourg and Oslo and of the CNRS INSU ALEAS program. We thank the Research Council of Norway through its Centres of Excellence funding scheme, project number 262644. We are also thankful for the support of the Russian Government, through its grant number 14.W03.31.0002. Readers are most welcome to contact the authors for discussion.

\FloatBarrier
\appendix

\section{Method for the measure of crack velocity versus energy release rate in PMMA\label{method}}

\begin{figure}[t]
  \includegraphics[width=1\linewidth]{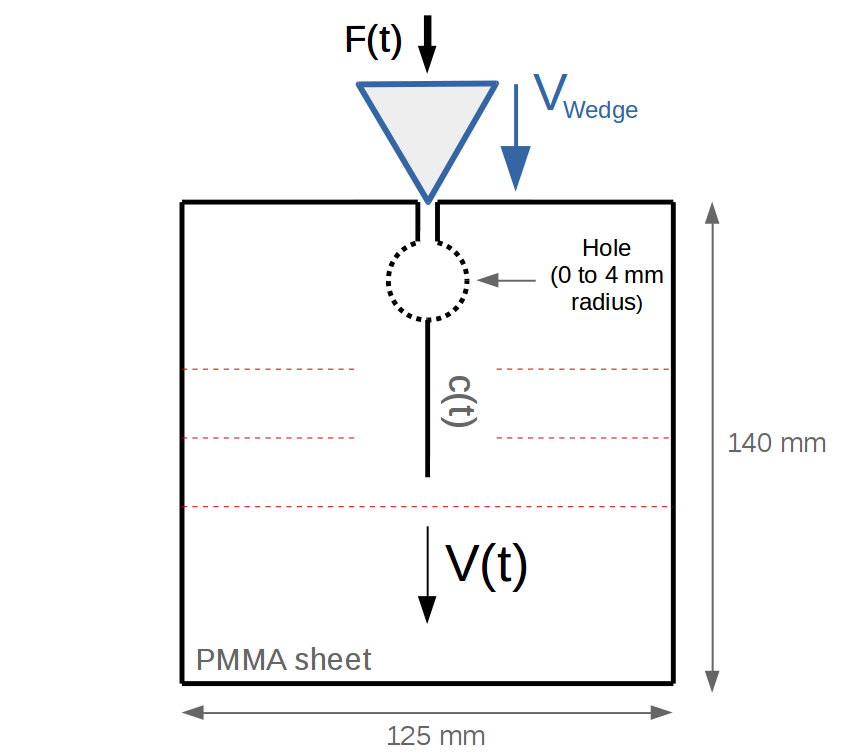}
  \caption{Schematic of the experimental set-up used to measure the crack energy release rate $G$ and its corresponding propagation velocity $V(t)$ in PMMA. See Refs.\,\cite{PMMA1,PMA3ss} for details. The hole is used to store some potential energy in the PMMA sheet for fast propagation experiments and is replaced by only a seed crack for slow ones. The dashed horizontal lines represent conductive metallic lines deposited onto the sample to measure the fast crack velocity with an oscilloscope.}
  \label{fig:setup}
\end{figure}

As part of the PMMA data was not published before (i.e., the slow propagation branch), we describe, in this section, the method that was used to acquire it.\\
Wedge splitting fracture tests are used to measure both the slow and fast $V(G)$ branches in PMMA\,\cite{PMMA1,PMA3ss}, whose geometry is shown in Fig.\,\ref{fig:setup}. Rectangular plates of size $140$\,mm\,$\times\,125$\,mm\,$\times\,15$\,mm are first machined from a plate of moulded PMMA (Perspex\textsuperscript{\textregistered}). A $25$\,mm\,$\times\,25$\,mm notch is subsequently cut out on one of the two lateral edges and a $8$-mm-long $800$-$\upmu$m-thick groove is finally introduced in the middle of the notch with a diamond saw.\\
To grow slow cracks, an additional seed crack ($\sim 2\,$mm-long) is added at the end of the groove via a razor blade. This crack is loaded in tension by pushing a steel wedge (semi-angle of $15^\circ$) in the notch. Two steel blocks equipped with rollers are placed in between the wedge and the specimen notch to limit the parasitic mechanical dissipation through plastic deformations or friction at loading contacts. As a result, the vicinity of the crack tip can be assumed to be the sole dissipation source for mechanical energy in the system. The wedge speed is first set to $1.6$\,µm\,s\textsuperscript{-1}. The force $F$, applied by the wedge to the specimen, increases linearly with time up to a point $F_c$ above which the seed crack starts to propagate. Above this point, $F$ decreases with time. We let the crack propagate over a distance of about $10\,$mm. This ensures reproducible initial conditions with a long-enough well-defined sharp seed crack. The specimen is then unloaded (unloading wedge speed: $16\,$µm\,s\textsuperscript{-1}). The specimen is then loaded again at a constant prescribed wedge speed $V_\text{wedge}$, which has been varied from $1.6\,$µm\,s\textsuperscript{-1} to $1.2\,$mm\,s\textsuperscript{-1}.
\\During each fracture test, the force $F(t)$ is monitored in real-time via a cell force mounted on the system (S-type Vishay load cell). A camera (USB2 uEye from IDS) is also used to image crack propagation at the specimen surface (space and time accuracy of $125\,\upmu$m and $0.1\,$s). A coarse approximation of the crack speed can be obtained by differentiating the position of the crack tip observed on the successive images. However, a more accurate signal $V(t)$ is obtained from the force signal (see Ref.\,\cite{Bares14} for details on the method). Indeed, in a linear elastic isotropic material like PMMA, the specimen stiffness $k(t)=F(t)/(V_\text{wedge} t)$ is a continuous decreasing function of the crack length, $c(t)$, that is set by the specimen geometry only. This function has been obtained using finite element calculations on the exact experimental geometry (Cast3M software, 2D simulation assuming plane stress conditions); it was checked that the obtained $k$ versus $c$ curve coincides with the experimental curves obtained by plotting $k(t)$ as a function of the crack length measured by the camera. The idea is then to use this curve $k(c)$, and the corresponding inverse function $k^{-1}$, to infer the time evolution of crack length $c(t)$ from the signal $F(t)$: $c(t) = k^{-1} [F(t)/(V_\text{wedge} t)]$. Time derivation of the so-obtained $c(t)$ provides a signal $V(t)$ about 50 times less noisy than that directly obtained from the camera images. The knowledge of $c(t)$ and $F(t)$ also allows determining the time evolution of the energy release rate, $G(t)$. Indeed, the total amount of mechanical energy provided to the specimen is $F^2(t)/[2 k(c(t))]$. Differentiating this stored energy with respect to $c$ directly provides $G(t)$. The slow branch of Fig.\,\ref{fig:altuglas} then provides the observed $V(t)$ as a function of $G(t)$. The results from twelve fracture experiments are gathered in this branch and differ by their $V_\text{wedge}$ value.\\
To grow fast cracks and measure $V(G)$ in the fast stable phase, the seed crack has been replaced by a hole of tunable radius ($1$ to $4\,$mm) drilled at the end of the groove\,\cite{PMMA1}. This delays fracture and increases the potential energy stored in the specimen at the initiation of crack growth. The time evolution of $V(t)$ is measured by monitoring, via an oscilloscope, the successive rupture of parallel $500$-$\upmu$m-large metallic lines (chromium/gold) deposited on the surface. That of the stress intensity factor $K$ is obtained via finite element analysis (see Ref.\,\cite{PMMA1} for details). The time evolution of the mechanical energy release rate is then deduced: $G=K^2/E$ where the Young modulus $E$ in the studied PMMA have been measured to be $E=2.8\,$GPa.
The fast branch of Fig.\,\ref{fig:altuglas} then provides the observed $V(t)$ as a function of $G(t)$. The results from five fracture experiments are gathered in this branch; they differ by the amount of stored elastic energy at crack growth initiation.

\section{Parameters sensitivity\label{fit}}

\begin{figure}
  \includegraphics[width=0.95\linewidth]{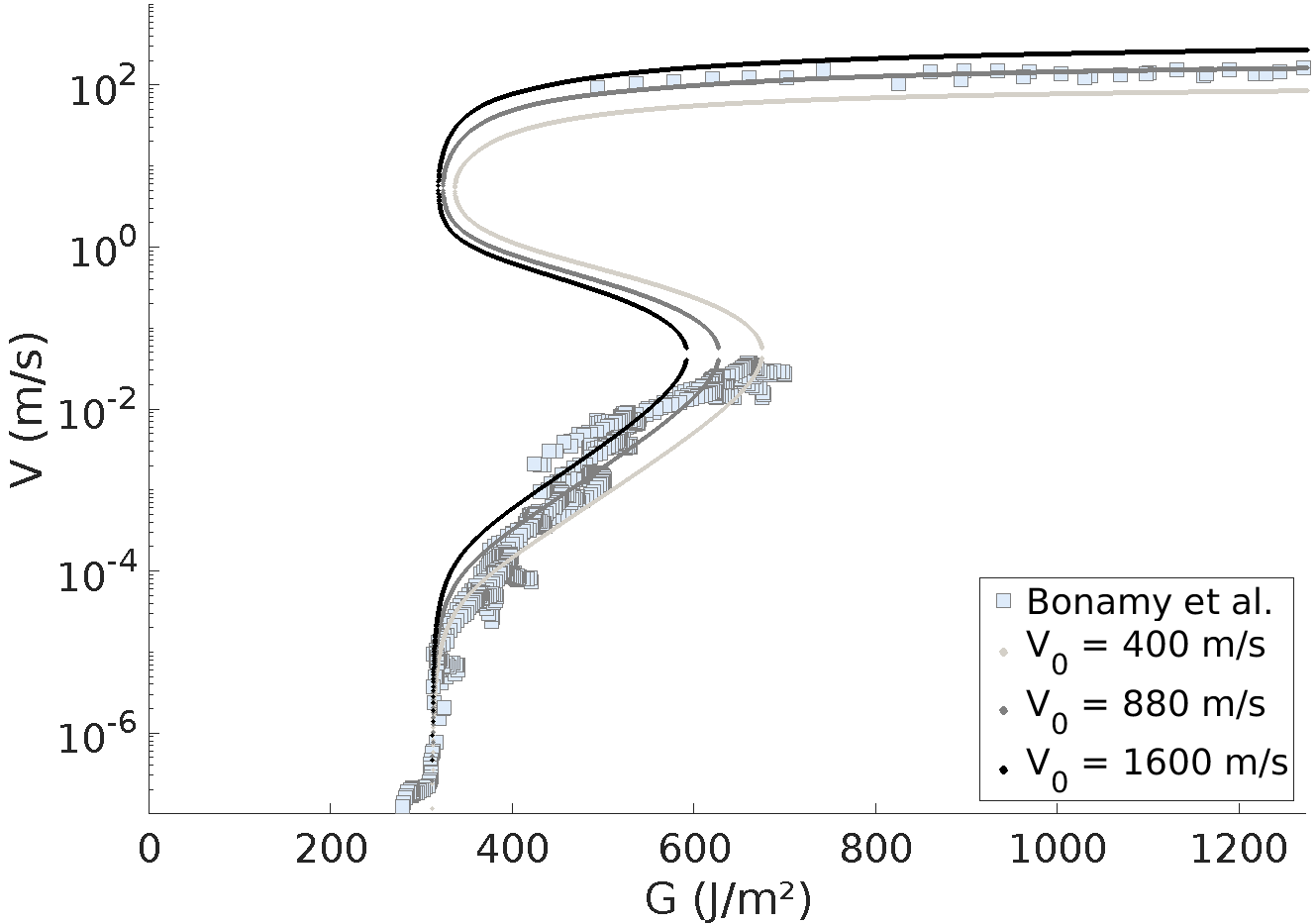}
  \caption{Effect of varying the nominal velocity, $V_0$, on the fit to the PMMA data. The propagation velocity is roughly proportional to $V_0$, but also modifies the positions of the phase transitions.}
  \label{fig:v0}
\end{figure}
\begin{figure}
  \includegraphics[width=0.95\linewidth]{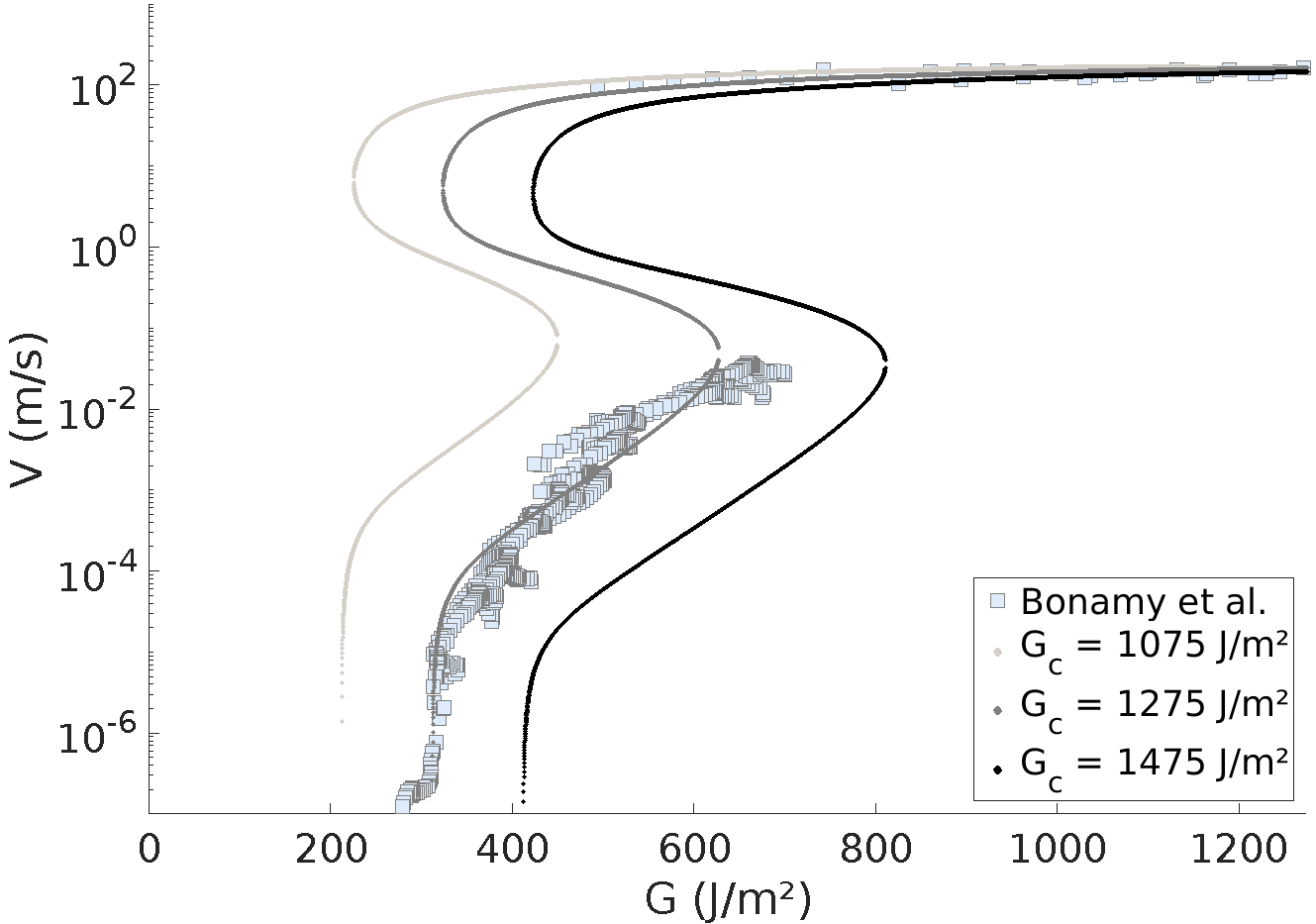}
  \caption{Effect of varying the breaking energy barrier, $G_c$, on the fit to the PMMA data. At a given load $G$, the higher $G_c$, the slower the crack. The transitions between the three propagation modes (fast, slow, and dominated by healing) are also affected: a medium with a stronger barrier needs a heavier load to transit to a weaker state.}
  \label{fig:Gc}
\end{figure}
\begin{figure}
  \includegraphics[width=0.95\linewidth]{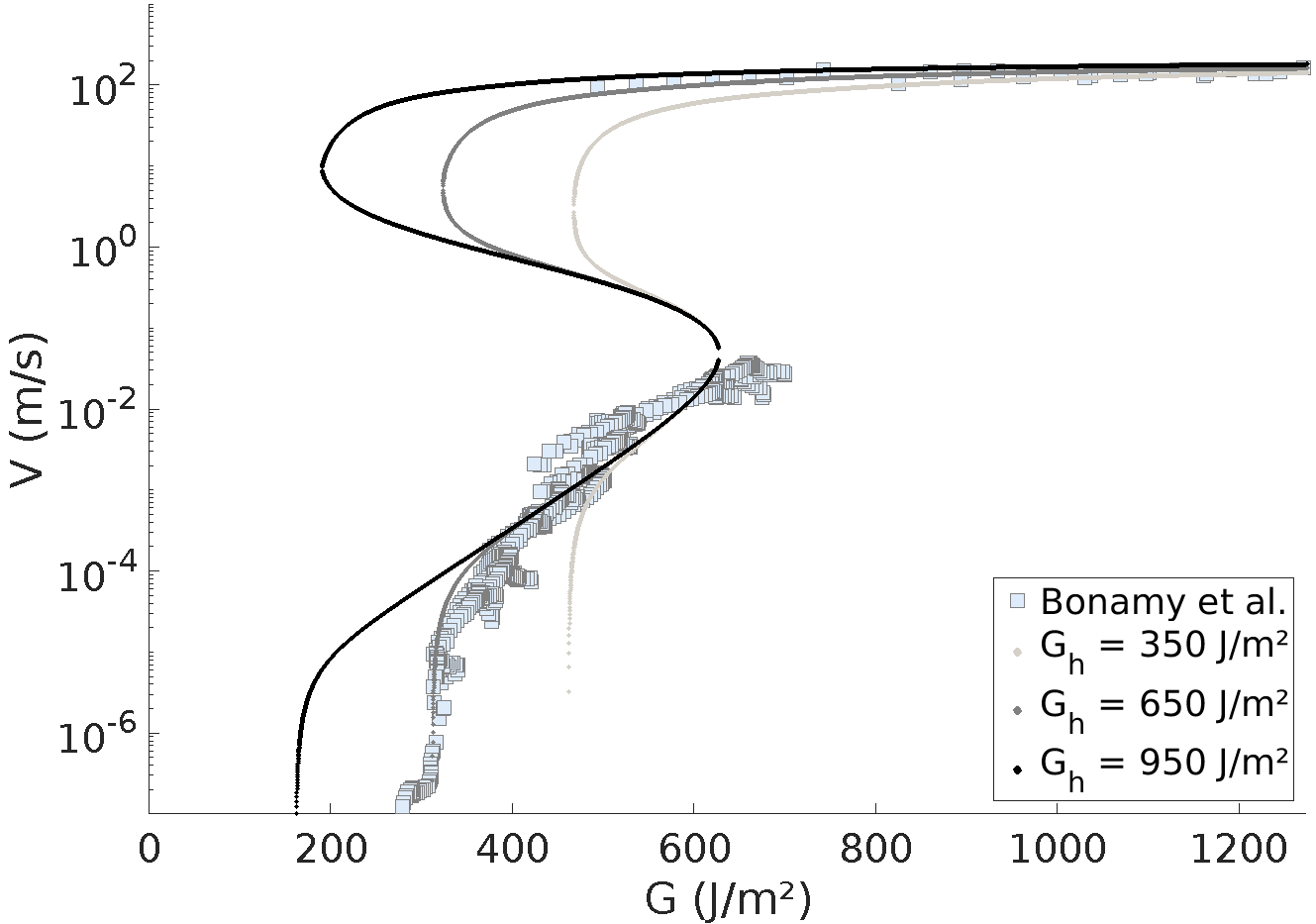}
  \caption{Effect of varying the healing energy barrier, $G_h$, on the fit to the PMMA data. A crack that heals more easily needs a higher load to actually propagate forward or to stay in the high velocity regime.}
  \label{fig:Gh}
\end{figure}
\begin{figure}
  \includegraphics[width=0.95\linewidth]{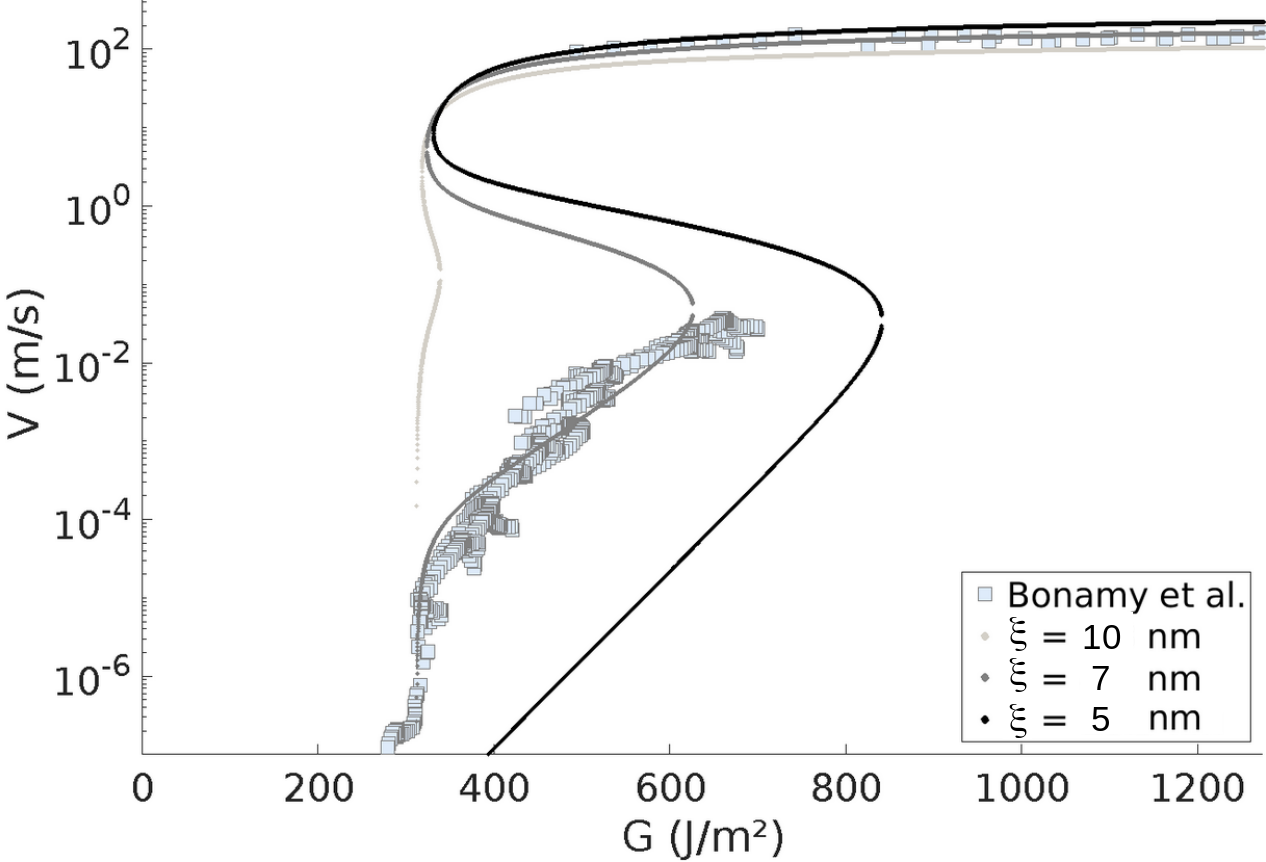}
  \caption{Effect of varying the stress cut-off scale $\xi$, on the fit to the PMMA data. $\xi$ mainly controls the slope and the intercept of the low velocity branch. A small change in $\xi$ significantly modifies this branch as well as the threshold to the fast regime.}
  \label{fig:xi}
\end{figure}
\begin{figure}
  \includegraphics[width=0.95\linewidth]{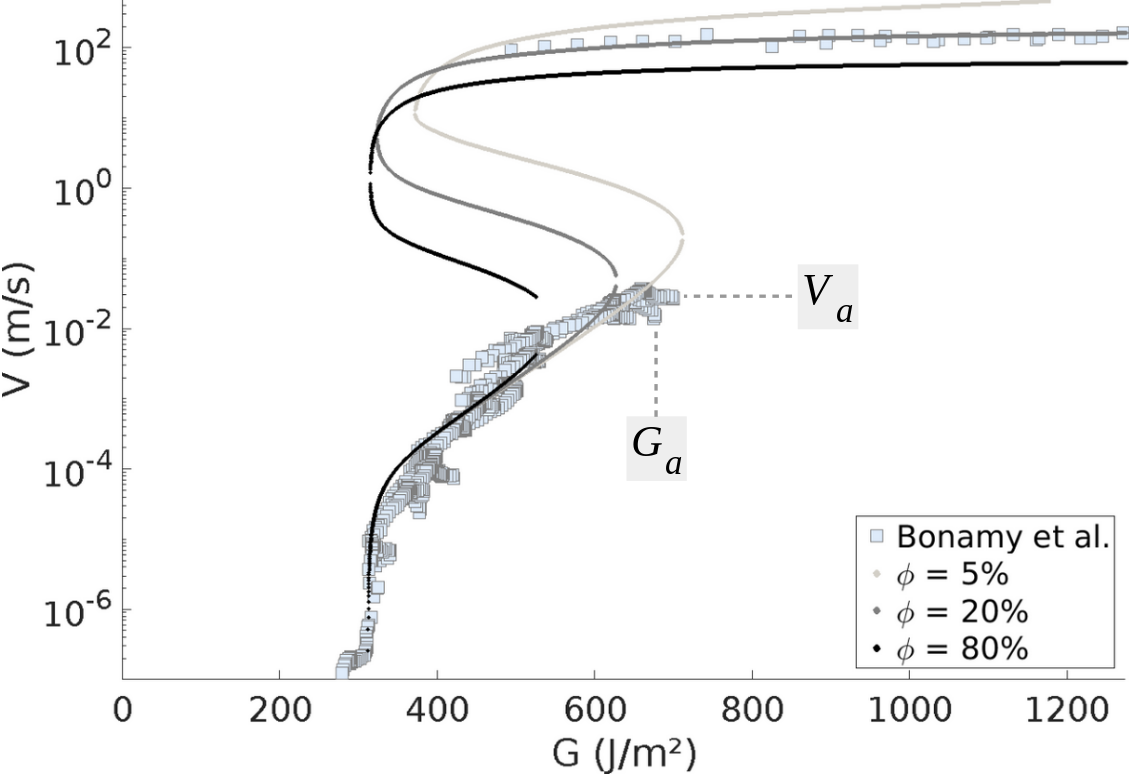}
  \caption{Effect of varying the ratio of energy converted to heat, $\phi$, on the fit to the PMMA data. The maximum velocity increases with $\phi$ as the tip temperature is higher. The threshold from the slow to the fast branch (i.e., the ($G_a$, $V_a$) point) shifts towards a lower $G$ as a lighter load is required for the temperature to significantly deviate from $T_0$.}
  \label{fig:phi}
\end{figure}
\begin{figure}
  \includegraphics[width=0.95\linewidth]{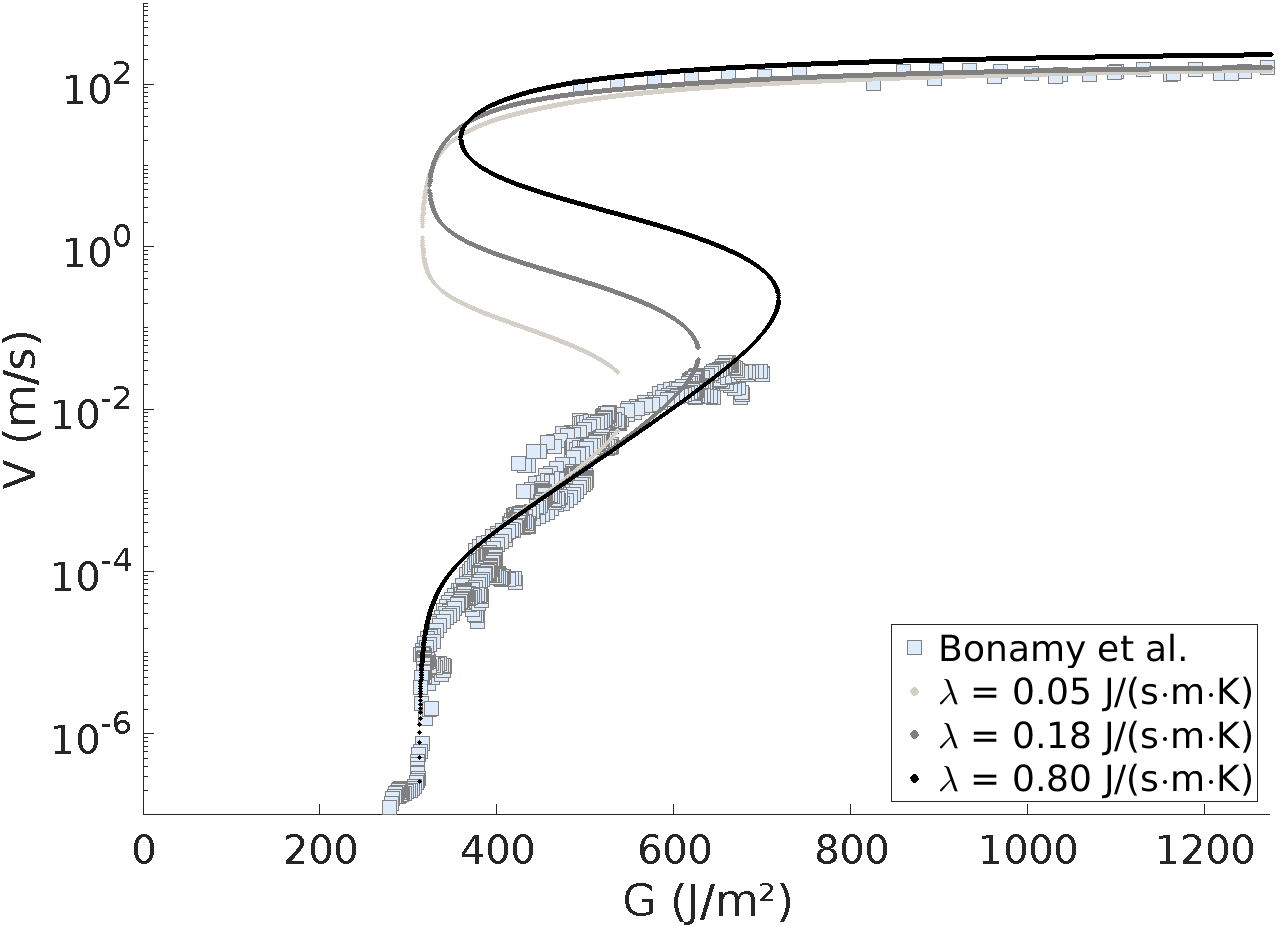}
  \caption{Effect of varying the thermal conductivity, $\lambda$, on the fit to the PMMA data. With a higher $\lambda$, the heat is better evacuated: the slow to fast branch threshold shifts towards higher $G$ and $V$. The fast regime is not very sensitive to $\lambda$, as $\Delta T$ is there constrained by $l$.}
  \label{fig:lambda}
\end{figure}
\begin{figure}
  \includegraphics[width=0.95\linewidth]{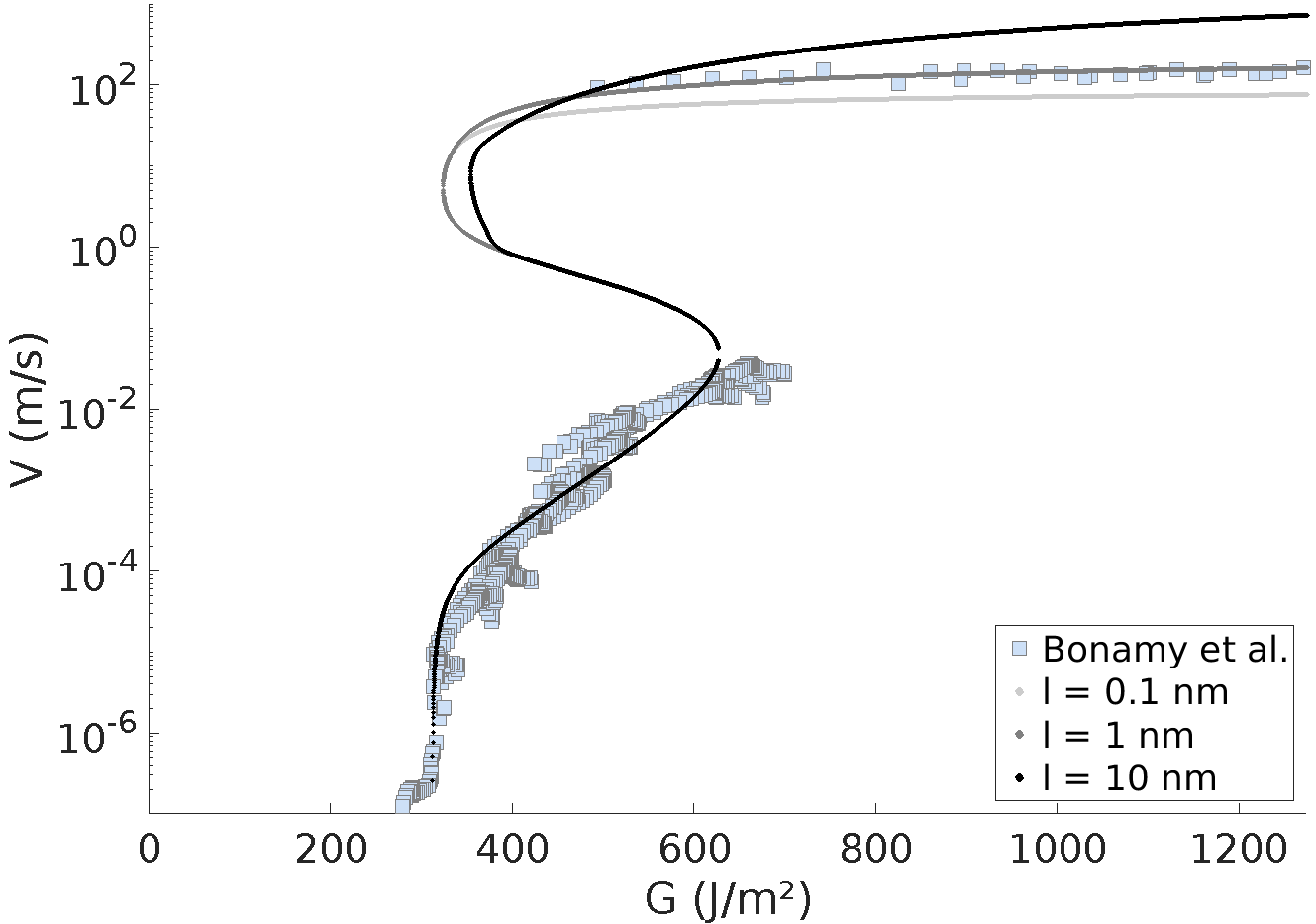}
  \caption{Effect of varying the heat production zone radius, $l$, on the fit to the PMMA data. $l$ mainly impacts the plot curvature on the high velocity branch. No effect is observed on the slow branch, as the thermal elevation there is constrained by the diffusion skin depth rather than by $l$ (see Eq.\,(\ref{slow})).}
  \label{fig:l}
\end{figure}
We here show, on the PMMA data, how varying the model parameters around their inferred values impacts the model fit, thus giving the reader a better feeling for their individual effect and sensitivity. In each of the figures \ref{fig:v0} to \ref{fig:l}, a unique parameter of the model varies while the others are kept to the exact values used for the fit presented in Fig.\,\ref{fig:altuglas}: $\xi = 56$\,nm, $V_0 = 880$\,m\,s\textsuperscript{-1}, $G_c = 1275$\,J\,m\textsuperscript{-2}, $G_h = 650$\,J\,m\textsuperscript{-2}, $\phi=20$\%, $\lambda=0.18$\,J\,s\textsuperscript{-1}\,m\textsuperscript{-1}\,K\textsuperscript{-1}, $C = 1.5\times10^{6}$\,J\,m\textsuperscript{-3}\,K\textsuperscript{-1}, $l=1$\,nm and $T_0 = 296$\,K. These seven plots show the fits up to the apparition of the secondary micro-cracks (see section\,\ref{micro} and Fig.\,\ref{fig:micro}), after which the model does not apply as such.
\\Naturally, some care should be taken when interpreting the inverted parameters (i.e., $\xi$, $G_c$, $G_h$, $\phi$ and $l$) beyond their actual orders of magnitude. For instance, $\xi$ and $G_c$ were fitted by a linear regression (i.e., Eq.\,(\ref{linear})) on the data which lies between $G=350$ and $G=700$\,J\,m\textsuperscript{-2} in Fig.\,\ref{fig:altuglas}, and the above values ($\xi=56$\,nm and $G_c=1275$\,J\,m\textsuperscript{-2}) were obtained with a coefficient of determination $R^2$ equal to $0.85$. Allowing $R^2$ to drop down to $0.75$ during this fit gives $\xi$ in a $30$ to $80$\,nm range and $G_c$ between $950$ and $1500$\,J\,m\textsuperscript{-2}. $G_h$ being directly deduced from $G_c$ (and from the vertical asymptote at low velocity in Fig.\,\ref{fig:altuglas}), the uncertainty on its value is comparable to that of $G_c$, that is, a few hundreds of joules per square meter. Let us now assess the accuracy of $\phi$. In the model, this parameters mainly controls which is the fastest point of the slow velocity branch (e.g., as shown in Fig.\,\ref{fig:phi}), after which cracks have to avalanche\,\cite{TVD1}. We then compare the experimental value for this particular point (obtained at $G=G_a$ and $V=V_a$, see Fig \ref{fig:phi}) to the model prediction of the same point. We quantify the error there as the euclidean distance between these points, in the sense of Eq.\,(\ref{error}). Such a relative, unitless, error minimizes to $10$\% for $\phi=0.25$ and, should we allow it to rise up to $30$\%, we obtain $\phi$ to be between $0.15$ and $0.30$. Finally, let us assess the accuracy of the inversion for the length scale of the heat production zone $l$. We vary $l$ and compute the same relative euclidean error in average over the fast velocity branch (i.e., the location where the model is mostly sensitive to $l$, see Fig.\,\ref{fig:l}). This error now minimizes to $5$\% for $l=1$\,nm and, letting it reach $30$\%, we infer $l$ to lie in a $0.1$\,nm to $2$\,nm range.\\
These uncertainties in the parameter inversion are somewhat high, but we here quantify an atomic scale process based on macroscopic measurements, so that this is not particularly surprising. One also needs to add up the experimental inaccuracy for $V$ and $G$ (see the data spread in Fig.\,\ref{fig:altuglas}), as well as the limitations of our very first order physical model, as discussed in section\,\ref{concl1}. Still, overall, the data is well explained over eight decades of velocities and with parameters that are in physically reasonable orders of magnitude.

\section{Data binning to compute a mean fit error\label{meanfit}}

\begin{figure}[ht]
  \includegraphics[width=0.9\linewidth]{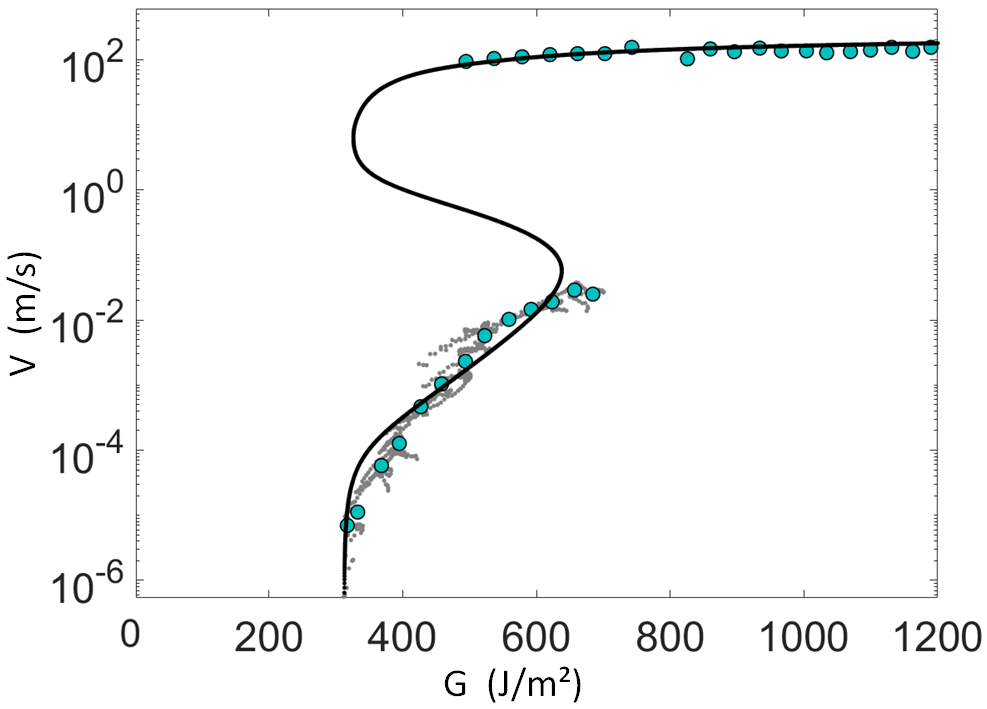}
  \caption{Under sampled PMMA data (blue circles), using a running average on $40$\,J\,m\textsuperscript{-2} bins. The dots are the original data points and the black line the fitted model. With this decimated data set, the mean fitting error is $\overline{\varepsilon} = 4\%$. The data is only shown to the onset of micro-cracking (i.e., see section\,\ref{micro}), beyond which the model does not apply as such.}
  \label{undersampl_altu}
\end{figure}

\begin{figure}[ht]
  \includegraphics[width=0.9\linewidth]{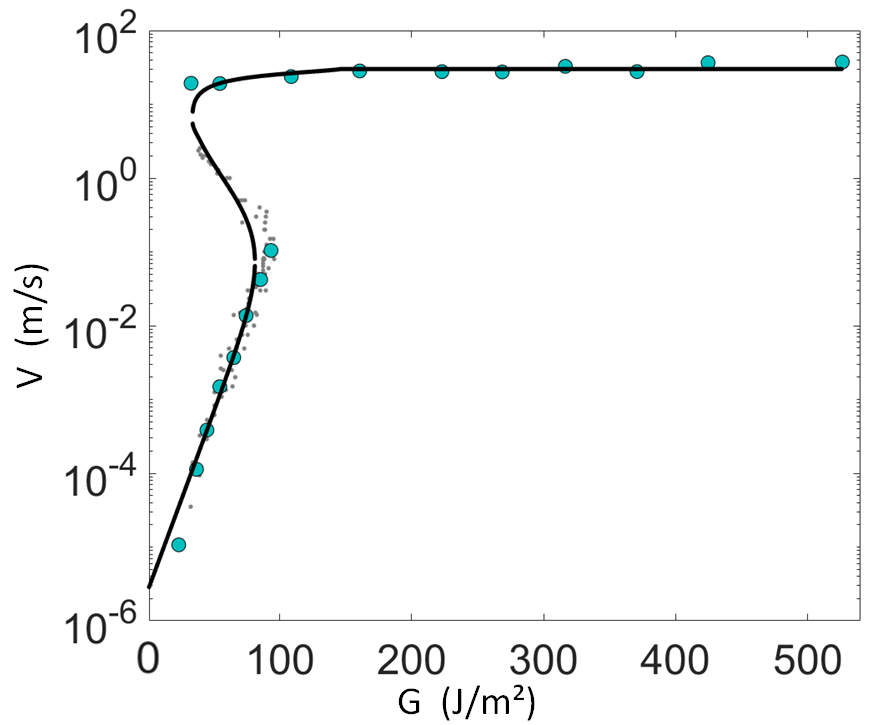}
  \caption{Under sampled PSA data (blue circles), using a running average on $10$\,J\,m\textsuperscript{-2} bins. The dots are the original data points and the black line the fitted model. With this decimated data set, the mean fitting error is $\overline{\varepsilon} = 5\%$.}
  \label{undersample_tape}
\end{figure}

To compute a mean fit error $\overline{\varepsilon} = \text{mean}_d(\varepsilon_d)$, where $\varepsilon_d$ is defined by Eq.\,(\ref{error}), we first binned the PMMA and PSA data points onto coarse bins using a running average on both stable branches, as explained in the core text (see section\,\ref{section:fits}). This was done to avoid the densely populated parts of the data sets to dominate on the value of the inferred $\overline{\varepsilon}$. Figures\,\ref{undersampl_altu} and\,\ref{undersample_tape} show the result of this data decimation.

\section{Healing processes in tape\label{ap_tape}}

We considered the healing processes to be negligible in order to describe the dynamics of unrolling tape, as no low velocity constant $G$ asymptote arising from crack healing displays in the ($G$, $V$) data (i.e., in Fig.\,\ref{fig:tape}). Such an absence would, however, also happen if $G_c$ was to be smaller than $G_h$, as the asymptote is obtained for $(G_c-C_h)/2$. Thus, the healing energy barrier could still be comparable to the breaking one, and so still significantly impact the high velocity propagation branch, when the crack tip is hot enough for healing to be non negligible (as predicted by Eq.\,(\ref{arrh1})). Of course, an accurate quantification of this effect suffers from the absence of the asymptote as it is the only good constraint for $G_h$. 
\begin{figure}[ht]
  \includegraphics[width=1\linewidth]{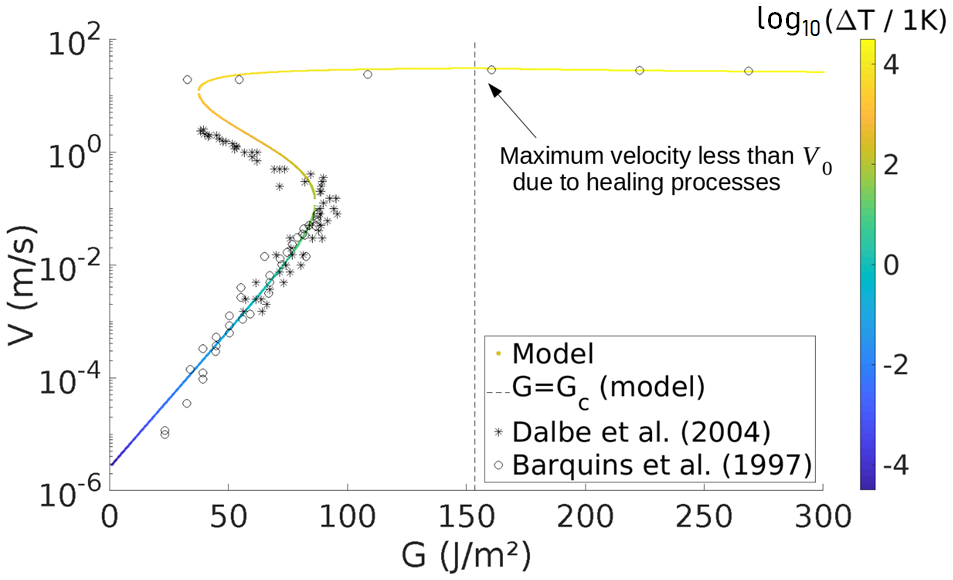}
  \caption{Fit of the Scotch\textsuperscript{\textregistered} 3M 600 data\,\cite{tape1, tape2} with a model including healing processes. The unstable (middle) branch of the model should not necessarily match the data point which are averaged $G$ and $V$ values for a front that stick-slips.}
  \label{fig:tape600_heal}
\end{figure}
\begin{figure}[ht]
  \includegraphics[width=1\linewidth]{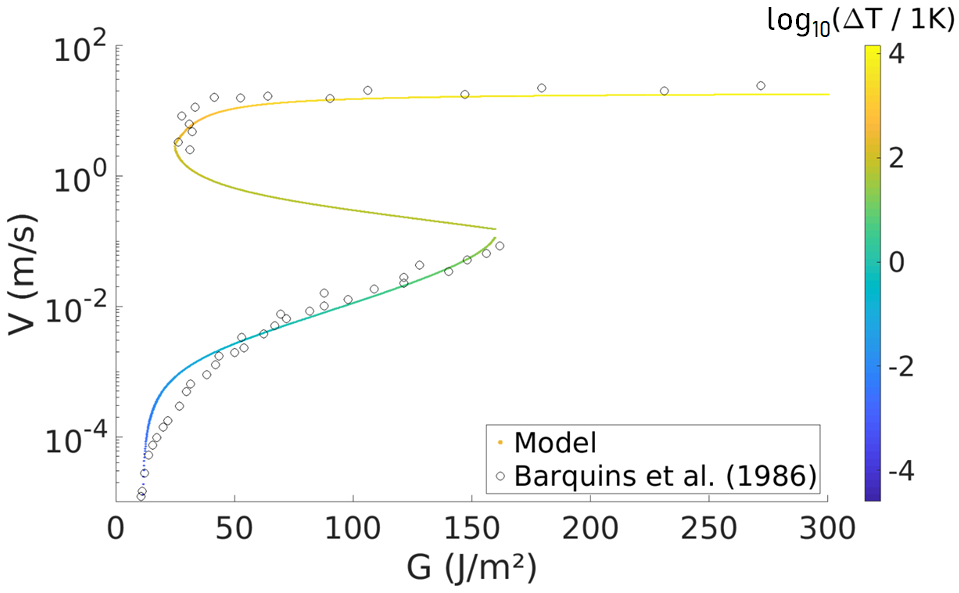}
  \caption{Fit of the peeling data for another roller tape: Scotch\textsuperscript{\textregistered} 3M 602 data\,\cite{tape602}. The lack of linearity at low velocity calls for healing processes in our model. Note also the curvature on the lower end of the high velocity branch, not present in the other data sets that we show but rather compatible with our proposed model.}
  \label{fig:tape602_heal}
\end{figure}
Figure \ref{fig:tape600_heal} shows for instance a model not disregarding healing, and compares it with the tape data. The match is there improved compared to the fit presented in Fig.\,\ref{fig:tape} as we have now an additional degree of freedom. The fit parameters in this figure are as follow: $\xi = 9$\,nm, $V_0 = 70$\,m\,s\textsuperscript{-1}, $G_c = 154$\,J\,m\textsuperscript{-2}, $G_h = 200$\,J\,m\textsuperscript{-2}, $\phi\sim 1$, $\lambda=0.2$\,J\,s\textsuperscript{-1}\,m\textsuperscript{-1}\,K\textsuperscript{-1}, $C = 10^{6}$\,J\,m\textsuperscript{-2}\,K\textsuperscript{-1}, $l=1$\,nm and $T_0 = 296$\,K.
\\Note that \citet{tape602}, who released part of the data presented in Fig.\,\ref{fig:tape600_heal}, also provided similar measurements for another type of roller tape, Scotch\textsuperscript{\textregistered} 3M 602 (see Ref.\,\cite{tape602}, in French). For this new medium, an asymptote does seem to display at low velocity on the ($G$, $V$) plot, calling for healing processes in our description, as shown in Fig.\,\ref{fig:tape602_heal}. We there propose a fit with the following parameters: $\xi = 40$\,nm, $V_0 = 200$\,m\,s\textsuperscript{-1}, $G_c = 500$\,J\,m\textsuperscript{-2}, $G_h = 480$\,J\,m\textsuperscript{-2}, $\phi=60$\%, $\lambda=0.3$\,J\,s\textsuperscript{-1}\,m\textsuperscript{-1}\,K\textsuperscript{-1}, $C = 10^{6}$\,J\,m\textsuperscript{-3}\,K\textsuperscript{-1}, $l=1$\,nm and\,$T_0 = 296$\,K.\\

\FloatBarrier
\bibliographystyle{unsrtnat}
\bibliography{scr.bib}

\end{document}